\documentclass[12pt]{article}

\usepackage{amssymb}
\usepackage{amsmath} 
\usepackage{amstext}
\usepackage{amsthm}
\usepackage{epsfig}
\usepackage{graphics}
\usepackage{bm}
\usepackage{color,verbatim}

\def\boxit#1{\vbox{\hrule\hbox{\vrule\kern6pt
          \vbox{\kern6pt#1\kern6pt}\kern6pt\vrule}\hrule}}

\allowdisplaybreaks

\renewcommand{\hat}{\widehat}

\numberwithin{equation}{section}

\def\nh{\noindent\hangindent=1.5truecm\hangafter=1}
\def\bs{\bigskip}

\def\ni{\noindent}

\def\a{\alpha}
\def\ab{\allowbreak}
\def\as{^*}

\def\bigmi{\,\big|\,}
\def\bX{{\bar X}}
\def\cD{{\cal D}}
\def\cE{{\cal E}}
\def\cF{{\cal F}}
\def\cG{{\cal G}}
\def\cS{{\cal S}}
\def\cX{{\cal X}}
\def\const{{\rm const.}\,}

\def\De{\Delta}
\def\ep{\epsilon}
\newcommand{\eps}{\epsilon}

\def\etal{{\sl et al.}~}
\def\etalc{{\sl et al.},~}

\def\ga{\gamma}
\def\hatt{{\hat t}}

\def\hc{{{\rm hc}}}

\def\hG{{\widehat G}}
\def\hga{{\hat\gamma}}
\def\half{^{1/2}}

\def\j{^{(j)}}
\def\la{\lambda}

\def\mi{\,|\,}
\def\mo{^{-1}}
\def\mt{^{-2}}

\def\mhf{^{-1/2}}
\def\one{^{(1)}}
\def\onn{{\textstyle{1\over9}}}
\def\oon{{1\over n}}
\def\oqr{{\textstyle{1\over4}}}
\def\osx{{\textstyle{1\over6}}}
\def\otd{{\textstyle{{1\over3}}}}

\def\tqr{{\textstyle{3\over4}}}
\def\ttd{{\textstyle{{2\over3}}}}
\def\p{^{(p)}}
\def\ra{\to}
\def\rai{\to\infty}
\def\si{\sigma}
\def\sumi{\sum_i\,}
\def\sumion{\sum_{i=1}^n\,}

\def\sumjop{\sum_{j=1}^p\,}
\def\tcD{{\widetilde\cD}}
\def\tcE{{\widetilde\cE}}
\def\Th{\Theta}
\def\th{\theta}

\def\thf{{\textstyle{1\over2}}}

\def\var{{\rm var}\,}
\def\ze{\zeta}
\def\goto{\rightarrow}

\newtheoremstyle{break}
  {9pt}
  {12pt}
  {\upshape}
  {}
  {\bfseries}
  {.}
  {\newline}
  {}

\newtheorem{theo}{Theorem}
\newtheorem{theoapp}{Theorem}

\theoremstyle{break}

\newtheorem{definition}{Definition}[section]

\newcommand{\bed}{\begin{definition}}
\newcommand{\eed}{\end{definition}}

\theoremstyle{theorem}

\renewcommand{\baselinestretch}{1.4}

\makeatletter
\def\singlespace{\deltaf\baselinestretch{1}\@normalsize}

\begin{document}

\begin{center}
{\Large \bf Robustness and accuracy of methods for high dimensional data analysis based on Student's $t$ statistic}
\end{center}

\vspace{0.3cm}
\begin{center}
Aurore Delaigle$^{1,2}$\quad Peter Hall$^{1,3}$\quad  and \quad Jiashun Jin$^4$\\

\begin{small}
{$^1$ Department of Mathematics and Statistics, University of Melbourne,  VIC 3010, Australia.}\\
{$^2$ Department of Mathematics, University of Bristol, Bristol BS8 1TW, UK.}\\
{$^3$ Department of Statistics, University of California at Davis, Davis, CA 95616, USA.}\\
{$^4$ Department of Statistics, Carnegie Mellon University, Pittsburgh, PA 15213, USA}
\end{small}
\end{center}
\vspace{1.3cm}

\setlength{\baselineskip}{0.8\baselineskip}

\noindent {\bf Abstract:} Student's $t$ statistic is finding applications today that were never envisaged when it was introduced more than a century ago.  Many of these applications rely on properties, for example robustness against heavy tailed sampling distributions, that were not explicitly considered until relatively recently.  In this paper we explore these features of the $t$ statistic in the context of its application to very high dimensional problems, including feature selection and ranking, highly multiple hypothesis testing, and sparse, high dimensional signal detection.  Robustness properties of the $t$-ratio are highlighted, and it is established that those properties are preserved under applications of the bootstrap.  In particular, bootstrap methods correct for skewness, and therefore lead to second-order accuracy, even in the extreme tails.  Indeed, it is shown that the bootstrap, and also the more popular but less accurate $t$-distribution and normal approximations, are more effective in the tails than towards the middle of the distribution.  These properties motivate new methods, for example bootstrap-based techniques for signal detection, that confine attention to the significant tail of a statistic.   

\vspace{1.3cm}

 \\   

\ni{\bf AMS SUBJECT CLASSIFICATION.}  \ Primary 62G32; Secondary 62F30, 62G90, 62H15, 62H30\\

\noindent {\bf Short title.}  Student's $t$ statistic

\newpage

\setlength{\baselineskip}{1.25\baselineskip}

\input epsf
\input colordvi.tex

\section{Introduction} \label{1}

\ni Modern high-throughput devices generate data in abundance.  Gene microarrays comprise an iconic example; there, each subject is automatically measured on thousands or tens of thousands of standard features.  What has not changed, however, is the difficulty of recruiting new subjects, with the number of the latter remaining in the tens or low hundreds.  This is the context of so-called ``$p\gg n$ problems,'' where $p$ denotes the number of features, or the dimension, and $n$ is the number of subjects, or the sample size.  

For each feature the measurements across different subjects comprise samples from potentially different underlying distributions, and can have quite different scales and be highly skewed and heavy tailed.  In order to standardise for scale, a conventional approach today is to use $t$-statistics, which, by virtue of the central limit theorem, are approximately normally distributed when $n$ is large.  W.~S.~Gosset, when he introduced the Studentised $t$-statistic more than a century ago (Student, 1908), saw that quantity as having principally the virtue of scale invariance.  In more recent times, however, other noteworthy advantages of Studentising have been discovered.  In particular, the $t$ statistic's high degree of robustness against heavy-tailed data has been quantified.  For example, Gin\'e, G\"otze and Mason (1997) have shown that a necessary and sufficient condition for the Studentised mean to have a limiting standard normal distribution is that the sampled distribution lie in the domain of attraction of the normal law.  This condition does not require the sampled data to have finite variance.  Moreover, the rate of convergence of the Studentised mean to normality is strictly faster than that for the conventional mean, normalised by its theoretical (rather than empirical) standard deviation, in cases where the second moment is only just finite (Hall and Wang, 2004).  Contrary to the case of the conventional mean, its Studentised form admits accurate large deviation approximations in heavy-tailed cases where the sampling distribution has only a small number of finite moments (Shao,~1999).  

All these properties are direct consequences of the advantages conferred by dividing the sample mean, $\bX$, by the sample standard deviation,~$S$.  Erratic fluctuations in $\bX$ tend to be cancelled, or at least dampened, by those of $S$, much more so than if $S$ were replaced by the true standard deviation of the population from which the data were drawn. 

The robustness of the $t$-statistic is particularly useful in high dimensional data analysis, where the signal of interest is frequently found to be sparse.  For any given problem (e.g.~classification, prediction, multiple testing), only a small fraction of the automatically measured features are relevant.  However the locations of the useful features are unknown,  and we must separate them empirically from an overwhelmingly large number of more useless ones.  Sparsity gives rise to a shift of interest away from problems involving vectors of conventional size to those involving high dimensional data.  

As a result, a careful study of moderate and large deviations of the Studentised ratio is indispensable to understanding even common procedures for analysing high dimensional data, such as ranking methods based on $t$-statistics, or their applications to highly multiple hypothesis testing.  See, for example, Benjamini and Hochberg (1995), Pigeot (2000), Finner and Roters (2002), Kesselman \etal(2002), Dudoit \etal(2003), Bernhard \etal(2004), Genovese and Wasserman (2004), Lehmann \etal(2005), Donoho and Jin (2006), Sarkar (2006), Jin and Cai (2007),  Wu (2008),  Cai and Jin (2010) and Kulinskaya (2009).  The same issues arise in the case of methods for signal detection, for example those based on Student's $t$ versions of higher criticism; see Donoho and Jin (2004), Jin (2007) and Delaigle and Hall~(2009).  Work in the context of multiple hypothesis testing includes that of Lang and Secic (1997, p.~63), Tamhane and Dunnett (1999), Takada \etal(2001), David \etal(2005), Fan \etal(2007) and Clarke and Hall (2009). 

In the present paper we explore moderate and large deviations of the Studentised ratio in a variety of high dimensional settings. Our results reveal several advantages of Studentising.  We show that the bootstrap can be particularly effective in relieving skewness in the extreme tails.  Attractive properties of the bootstrap for multiple hypothesis testing were apparently first noted by Hall (1990), although in the case of the mean rather than its Studentised form. 

Section~\ref{sec:advantages} draws together several known results in the literature in order to demonstrate the robustness of the $t$ ratio in the context of high level exceedences.  Sections~\ref{sec:boot} and~\ref{sec:nonzero} show that, even for extreme values of the $t$ ratio, the bootstrap captures particularly well the influence that departure from normality has on tail probabilities.  We treat cases where the probability of exceedence is either polynomially or exponentially small.  Section~\ref{sec:multi} shows how these properties can be applied to high dimensional problems, involving potential exceedences of high levels by many different feature components.  One example of this type is the use of $t$-ratios to implement higher criticism methods, including their application to classification problems.  This type of methodology is taken up in section~\ref{sec:HC}.  The conclusions drawn in sections~2 and 3 are illustrated numerically in section~\ref{sec:numerical}, the underpinning theoretical arguments are summarised in section~5, and detailed arguments are given by Delaigle \etal(2010).

\section{Main conclusions and theoretical properties}

\subsection{Advantages and drawbacks of studentising in the normal approximation}\label{sec:advantages}
Let $X_1,X_2,\ldots$ denote independent univariate random variables all distributed as $X$, with unit variance and zero mean, and suppose we want to test $H_0:\mu=0$ against $H_1:\mu>0$.  Two common test statistics for this problem are the standardised mean $Z_0$ and the Studentised mean $T_0$, defined by $Z_0=n\half\,\bX$ and $T_0=Z_0/S$ where 
\begin{align}
\bX=\oon\,\sumion X_i\,,\quad 
S^2=\oon\,\sumion(X_i-\bX)^2
\label{(2.1)}
\end{align}
denote the sample mean and sample variance, respectively, computed from the dataset $X_1,\ldots,X_n$.

In practice, experience with the context often suggests the standardisation that defines~$Z_0$. Although both $Z_0$ and $T_0$ are asymptotically normally distributed, dividing by the sample standard deviation introduces a degree of extra noise which can make itself felt in terms of greater impact of skewness.  However, we shall show that, compared to the normal approximation to the distribution of $Z_0$, the normal approximation to the distribution of $T_0$  is valid under much less restrictive conditions on the tails of the distribution of $X$. 

These properties will be established by exploring the relative accuracies of normal approximations to the probabilities $P(Z_0>x)$ and $P(T_0>x)$, as $x$ increases, and the conditions for validity of those approximations.  This approach reflects important applications in problems such as multiple hypothesis testing, and classification or ranking involving high dimensional data, since there it is necessary to assess the relevance, or statistical significance, of large values of sample means.

We start by showing that the normal approximation is substantially more robust for $T_0$ than it is for $Z_0$. To derive the results, note that if 
\begin{align}
E|X|^3<\infty\label{(2.2)}
\end{align}
then the normal approximation to the probability $P(T_0>x)$ is accurate, in relative terms, for $x$ almost as large as $n^{1/6}$.  In particular, $P(T_0>x)/\{1-\Phi(x)\}\ra1$ as $n\rai$, uniformly in values of $x$ that satisfy $0<x\leq\ep\,n^{1/6}$, for any positive sequence $\ep$ that converges to zero (Shao, 1999).  This level of accuracy applies also to the normal approximation to the distribution of the nonstudentised mean, $\bX$, except that we must impose a condition much more severe than~\eqref{(2.2)}.  In particular, $P(Z_0>x)/\{1-\Phi(x)\}\ra1$, uniformly in $0<x\leq n^{(1/6)-\eta}$, for each fixed $\eta>0$, if and only~if 
\begin{align}
E\big\{\exp\big(|X|^c\big)\big\}<\infty\quad
\hbox{for all}\quad c\in\big(0,\thf\big)\,;\label{(2.3)}
\end{align}
see Linnik~(1961).  Condition \eqref{(2.3)}, which requires exponentially light tails and implies that all moments of $X$ are finite, is much more severe than~\eqref{(2.2)}.

Although dividing by the sample standard deviation confers robustness, it also introduces a degree of extra noise. 
To quantify deleterious effects of Studentising we note that
\begin{align}
P(T_0>x)&=\{1-\Phi(x)\}\,\big\{1-n\mhf\,\otd\,x^3\,\ga+o\big(n\mhf\,x^3\big)\big\}\,,\label{(2.4)}\\
P(Z_0>x)&=\{1-\Phi(x)\}\,\big\{1+n\mhf\,\osx\,x^3\,\ga+o\big(n\mhf\,x^3\big)\big\}\,,\label{(2.5)}
\end{align}
uniformly in $x$ satisfying $\lambda_n\leq x\leq n^{1/6}\lambda_n$, for a sequence $\lambda_n\to\infty$, and where  $\Phi$ is the standard normal distribution function and $\ga=E(X^3)$ (Shao, 1999; Petrov, 1975, Chap.~8).  (Property \eqref{(2.2)} is sufficient for \eqref{(2.4)} if $x\rai$ and $n\mhf\,x^3\ra0$ as $n\rai$, and \eqref{(2.5)} holds, for the same range of values of $x$, provided that, for some $u>0$, $E\{\exp(u\,|X|)\}<\infty$.)  Thus it can be seen that, if $\ga\neq0$ and $n\mhf\,x^3$ is small, the relative error of the normal approximation to the distribution of $T_0$ is approximately twice that of the approximation to the distribution of~$Z_0$.

Of course, Student's $t$ distribution with $n$ or $n-1$ degrees of freedom is identical to the distribution of $T_0$ when $X$ is normal N$(0,\si^2)$, and therefore relates to the case of zero skewness. Taking $\gamma=0$ in \eqref{(2.4)} we see that, when $T_0$ has Student's $t$ distribution with $n$ or $n-1$ degrees of freedom, we have $P(T_0>x)=\{1-\Phi(x)\}\,\big\{1+o\big(n\mhf\,x^3\big)\big\}$. It can be deduced that the results derived in \eqref{(2.4)} and \eqref{(2.5)} continue to hold if we replace the role of the normal distribution by that of Student's $t$ distribution with $n$ or $n-1$ degrees of freedom.  Similarly, the results on robustness hold if we replace the role of the normal distribution by that of Student's $t$ distribution. Thus, approximating the distributions of $T_0$ and $Z_0$ by that of a Student's $t$ distribution, as is sometimes done in practice, instead of that of a normal distribution, does not alter our conclusions.  In particular, even if we use the Student's $t$ distribution, $T_0$ is still more robust against heavy tailedness than $Z_0$, and in cases where the Student approximation is valid, this approximation is slightly more accurate for $Z_0$ than it is for $T_0$. 

\subsection{Correcting skewness using the bootstrap}\label{sec:boot}
The arguments in section \ref{sec:advantages} show clearly that $T_0$ is considerably more robust than $Z_0$ against heavy-tailed distributions, arguably making $T_0$ the test statistic of choice even if the population variance is known. However, as also shown in section \ref{sec:advantages}, this added robustness comes at the expense of a slight loss of accuracy in the approximation.  For example, in \eqref{(2.4)} and \eqref{(2.5)} the main errors that arise in normal (or Student's $t$) approximations to the distributions of $T_0$ are the result of uncorrected skewness.  In the present section we show that if we instead approximate the distribution of $T_0$ using the bootstrap then those errors can be quite successully removed. 
Similar arguments can be employed to show that a bootstrap approximation to the distribution of $Z_0$ is less affected by skewness than a normal approximation. However, as for the normal approximation, the latter bootstrap approximation is only valid if the distribution of $X$ is very light tailed. Therefore, even if we use the bootstrap approximation, $T_0$ remains the statistic of choice.

Let $\cX\as=\{X_1\as,\ldots,X_n\as\}$ denote a resample drawn by sampling randomly, with replacement, from $\cX=\{X_1,\ldots,X_n\}$, and put 
\begin{align}
\bX\as=\oon\,\sumion X_i\as\,,\quad
S\as{}^2=\oon\,\sumion(X_i\as-\bX\as)^2\,,\quad
T_0\as=n\half\,(\bX\as-\bX)/S\as\,.\label{(2.6)}
\end{align}
The bootstrap approximation to the distribution function $G(t)=P(T_0\leq t)$ is $\hG(t)=P(T_0\as\leq t\mi\cX)$, and the bootstrap approximation to the quantile $t_\a=(1-G)\mo(\a)$ is 
\begin{align}
\hatt_\a=\big(1-\hG\big)\mo(\a)\,.\label{(2.7)}
\end{align}
Theorem~1, below, addresses the effectiveness of these approximations for large values of~$x$.

As usual in hypothesis testing problems, to calculate the level of the test we take a generic variable that has the distribution of the test statistic and we calculate the probability that the generic variable is larger than the estimated $1-\alpha$ quantile. This generic variable is independent of the sample, and since the quantile $\hatt_\a$ of the bootstrap test is random and constructed from the sample then, to avoid confusion, we should arguably use different notations for $T_0$ and the generic variable. However, to simplify notation we keep using $T_0$ for a generic random variable distributed like $T_0$. This means that we write the level of the test as $P(T_0>\hatt_\a)$, but here $T_0$ denotes a generic random variable independent of the sample, whereas $\hatt_\a$ denotes the random variable defined at \eqref{(2.7)} and calculated from the sample. In particular, here $T_0$ is independent of $\hatt_\a$.

Define $z_\a=(1-\Phi)\mo(\a)$, and write $P_F$ for the probability measure when $\cX$ is drawn from the population with distribution function~$F$. Here we highlight the dependence of the probabilities on $F$ because we shall use the results in subsequent sections where a clear distinction of the distribution will be required.  

\begin{theo}\label{Theorem1}  For each $B>1$ and $D_1>0$ there exists $D_2>2$, increasing no faster than linearly in $D_1$ as the latter increases, such that 
\begin{align}
P_F(T_0>\hatt_\a)
=\a\,\Big[1+O\big\{(1+z_\a)\,n\mhf+(1+z_\a)^4\,n\mo\big\}\Big]
+O\big(n^{-D_1}\big)\label{(2.8)}
\end{align}
as $n\rai$, uniformly in all distributions $F$ of the random variable $X$ such that $E(|X|^{D_2})\leq B$, $E(X)=0$ and $E(X^2)=1$, and in all $\a$ satisfying $0\leq z_\a\leq B\,n^{1/4}$.   
\end{theo}

The assumption in Theorem~1 that $E(X^2)=1$ serves to determine scale, without which the additional condition $E(|X|^{D_2})\leq B$ would not be meaningful for the very large class of functions considered in the theorem.  The theorem can be deduced by taking $c=0$ in Theorem~B in section~5.1, and shows that using the bootstrap to approximate the distribution of $T_0$ removes the main effects of skewness.  To appreciate why, note that if we were to use the normal approximation to the distribution of $T_0$ we would obtain, instead of \eqref{(2.8)}, the following result, which can be deduced from Theorem~A in section~5.1 for each $B>1$ such that $E|X|^4<B$ and $0\leq z_\a\leq B\,n^{1/4}$: 
\begin{align}
P_F(T_0>z_\a)
=\a\,\exp\big(-n\mhf\,\otd\,z_\a^3\,\ga\big)\,
\Big[1+O\big\{(1+z_\a)\,n\mhf+(1+z_\a)^4\,n\mo\big\}\Big]\,.
\label{(2.9)}
\end{align}
Comparing \eqref{(2.8)} and \eqref{(2.9)} we see that the bootstrap approximation has removed the skewness term that describes first-order inaccuracies of the standard normal approximation.  

The size of the $O(n^{-D_1})$ remainder in \eqref{(2.8)} is important if we wish to use the bootstrap approximation in the context of detecting $p$ weak signals, or of hypothesis testing for a given level of family-wise error rate or false discovery rate among $p$ populations or features.  (Here and below it is convenient to take $p$ to be a function of $n$, which we treat as the main asymptotic parameter.)  In all these cases we generally wish to take $\a$ of size $p\mo$, in the sense that $p\a$ is bounded away from zero and infinity as $n\rai$. This property entails $z_\a=O\{(\log p)\half\}$, and therefore Theorem~1 implies that the tail condition $E(|X|^{D_2})<\infty$, for some $D_2>0$, is sufficient for it to be true that ``$P_F(T_0>\hatt_\a)/\a=1+o(1)$ for $p=o(n^{D_1})$ and uniformly in the class of distributions $F$ of $X$ for which $E(X)=0$, $E(X^2)=1$ and $E(|X|^{D_2})<\infty$.''  

On the other hand, if, as in Fan and Lv (2008), $p$ is exponentially large as a function of $n$, then we require a finite exponential moment of $X$. The following theorem addresses this case.  In the theorem, $D_2<2$ unless $D_1={3\over8}$, in which case $D_2=2$. The proof of the theorem is given in section \ref{sec:proofTheo2}.

\begin{theo}\label{Theorem2}  For each $B>1$ and $D_1\in(0,{3\over8}]$ there exists $D_2\in(0,2]$, increasing no faster than linearly in $D_1$ as the latter increases, such that 
\begin{align}
P_F(T_0>\hatt_\a)
=\a\,\Big[1+O\big\{(1+z_\a)\,n\mhf+(1+z_\a)^4\,n\mo\big\}\Big]
+O\big\{\exp\big(-n^{D_1}\big)\big\}\label{(2.10)}
\end{align}
as $n\rai$, uniformly in all distributions $F$ of the random variable $X$ such that $P(|X|>x)\leq C\,\exp(-x^{D_2})$ (where $C>0$), $E(X)=0$ and $E(X^2)=1$, and in all $\a$ satisfying $0\leq z_\a\leq B\,n^{1/4}$.   
\end{theo}

Theorem~2 allows us to repeat all the remarks made in connection with Theorem~1 but in the case where $p$ is exponentially large as a function of~$n$.  Of course, we need to assume that exponential moments of $X$ are finite, but in return we can control a variety of statistical methodologies, such as sparse signal recovery or false discovery rate, for an exponentially large number of potential signals or tests.  Distributions with finite exponential moments include exponential families and distributions of variables supported on a compact domain. Note that our condition is still less restrictive than assuming that the distribution is normal, as is done in many papers treating high dimensional problems, such as for example Fan and Lv (2008).

\subsection{Effect of a nonzero mean on the properties discussed in section~2.2}\label{sec:nonzero}
We have shown that, in a variety of problems, when making inference on a mean it is preferable to use the Studentised mean rather than the standardised mean. We have also shown that, when the skewness of the distribution of $X$ is non zero, the level of the test based on the Studentised mean is better approximated when using the bootstrap than when using a normal distribution. Our next task is to check that, when $H_0:\mu=0$ is not true, the probability of rejecting $H_0$ is not much affected by the bootstrap approximation. Our development is notationally simpler if we continue to assume that $E(X)=0$ and $\var(X)=1$, and consider the test $H_0: \mu=-cn^{-1/2}$ with $c> 0$ a scalar that potentially depends on $n$ but which does not converge to zero.
We define 
\begin{align}
Z_c=n\half\,\big(\bX+c\,n\mhf\big)\,,\quad
T_c=Z_c/S\,.\label{(2.11)}
\end{align}
Here we take $\mu$ of magnitude $n^{-1/2}$ because this represents the limiting case where inference is possible. Indeed, a population with mean of order $o(n^{-1/2})$ could not be distinguished from a population with mean zero. Thus we treat the statistically most challenging problem.

Our aim is to show that the probability $P_F(T_c>t_\a)$ is well approximated by $P_F(T_c>\hatt_\a)$, where $c>0$ and $\hatt_\a$ is given by \eqref{(2.7)}, and when $T_c$ and $\hatt_\a$ are computed from independent data. We claim that in this setting the results discussed in section~2.2 continue to hold. In particular, versions of \eqref{(2.8)} and \eqref{(2.10)} in the present setting~are:
\begin{align}
P_F(T_c>\hatt_\a) &= P_F(T_c>t_\alpha)\,
\Big[1+O\big\{(1+z_\a)\,n\mhf+(1+z_\a)^4\,n\mo\big\}\Big] + R\,,\label{(2.12)}
\end{align}
where $\ga=E(X^3)$ denotes skewness and the remainder term $R$ has either the form in \eqref{(2.8)} or that in \eqref{(2.10)}, depending on whether we assume existence of polynomial or exponential moments, respectively.  In particular, if we take $R=O(n^{-D_1})$ then \eqref{(2.12)} holds uniformly in all distributions $F$ of the random variable $X$ such that $E(|X|^{D_2})\leq B$, $E(X)=0$ and $E(X^2)=1$, and in all $\a$ satisfying $0\leq z_\a\leq B\,n^{1/4}$, provided that $D_2$ is sufficiently large; and in the same sense, but with $R=O\{\exp(-n^{D_1})\}$ where $D_1\in(0,{3\over8}]$, \eqref{(2.12)} holds if we replace the assumption $E(|X|^{D_2})\leq B$ by $P(|X|>x)\leq C\,\exp(-x^{D_2})$, provided that $D_2\in(0,2]$ is sufficiently large.  (We require $D_2=2$ only if $D_1={3\over8}$.)  Result \eqref{(2.12)} is derived in section~\ref{proof(2.12)}. 
Hence to first order, the probability of rejecting $H_0$ when $H_0$ is not true is not affected by the bootstrap approximation. In particular, to first order, skewness does not affect the approximation any more than it would if $H_0$ were true (compare with \eqref{(2.8)} and \eqref{(2.10)}).

An alternative form of  \eqref{(2.12)}, which is useful in applications (e.g.~in section \ref{sec:HC}), is to express the right hand side there more explicitly in terms of $\alpha$. This can be done if we note that, in view of Theorem~A in section~5.1, 
\begin{align}
P_F(T_c>t_\a)
&=\{1-\Phi(t_\a)\}\,
\exp\Big\{-n\mhf\,\osx\,\big(2\,t_\a^3-3\,c\,t_\a^2+c^3\big)\,\ga\Big\}\,
{1-\Phi(t_\a-c)\over1-\Phi(t_\a)}\notag\\
&\qquad\times
\Big[1+\th(c,n,t_\a)\,\Big\{(1+t_\a)\,n\mhf+(1+t_\a)^4\,n\mo\Big\}\Big]\notag\\
&=\a\,\exp\big\{n\mhf\,\osx\,c\,\big(3\,t_\a^2-c^2\big)\,\ga\big\}\,
{1-\Phi(t_\a-c)\over1-\Phi(t_\a)}\notag\\
&\qquad\times
\Big[1+\th_1(c,n,t_\a)\,\Big\{(1+t_\a)\,n\mhf+(1+t_\a)^4\,n\mo\Big\}\Big]\,,
\label{(5.18b)}
\end{align}
where $\th_1$ has the same interpretation as $\th$ in Theorem~A, and the last identity follows from the definition of~$t_\a$. Combining this property with \eqref{(2.12)} it can be shown that 
\begin{align}
P_F(T_c>\hatt_\a) &= \a\,\exp\big\{n\mhf\,\osx\,c\,\big(3\,t_\a^2-c^2\big)\,\ga\big\}\,{1-\Phi(t_\a-c)\over1-\Phi(t_\a)}\notag\\
&\qquad \times \Big[1+O\big\{(1+z_\a)\,n\mhf+(1+z_\a)^4\,n\mo\big\}\Big] + R,\label{(2.12b)}
\end{align}
where $R$ satisfies the properties given below \eqref{(2.12)}.

\subsection{Relationships among many events $T_c>\hatt_\a$}  \label{sec:multi}
So far we have treated only an individual event (i.e. a single univariate test), exploring its likelihood.  However, since our results for a single event apply uniformly over many choices of the distribution of $X$ then we can develop properties in the context of many events, and thus for simultaneous tests.  The simplest case is that where the values of $T_c$ are independent; that is, we observe $T_{c\j}\j$ for $1\leq j\leq p$, where $c\one,\ldots,c\p$ are constants and the random variables $T_{c\j}\j$ are, for different values $j$, computed from independent datasets.  We assume that $T_{c\j}\j$ is defined as at \eqref{(2.11)} but with $c=c\j$.  We could take the values of $n=n_j$ to depend on $j$, and in fact the theoretical discussion below remains valid provided that $C_1\,n\leq n_j\leq C_2\,n$, for positive constants $C_1$ and $C_2$, as $n$ increases.  (Recall that $n$ is the main asymptotic parameter, and $p$ is interpreted as a function of~$n$.)  As in the case of a single event, treated in Theorems~1 and~2, it is important that the $t$-statistic $T_{c\j}\j$ and the corresponding quantile estimator $\hatt_\a\j$ be independent for each~$j$.  However, as noted in section~2.2, this is not a problem since $T_{c\j}\j$ represents a generic random variable, and only $\hatt_\a\j$ is calculated from the sample.

Under the assumption that the variables $T_{c\j}\j$ and $\hatt_\a\j$, for $1\leq j\leq p$, are totally independent we can deduce from \eqref{(2.12)} that, uniformly in $z_\a$ satisfying $0\leq z_\a\leq B\,n^{1/4}$, 
\begin{align*}
P\big(T_{c\j}\j\bowtie_j z_\a\quad\hbox{for}\quad1\leq j\leq p\big)
&=\prod_{j=1}^p\,\psi_j\big(c\j\big)\,,
\end{align*}
where, for each $j$, $\bowtie_j$ denotes either $>$ or $\leq\,$, and 
\begin{align}
\psi_j(c)=\chi_j(c)&\equiv P\big(T_{c\j}\j\bowtie_j z_\a\big)
\exp\big\{n\mhf\,\osx\,c\,\big(3\,z_\a^2-c^2\big)\,\ga\j\big\}\notag\\
&\qquad\qquad\times
\Big[1+O\big\{(1+z_\a)\,n\mhf+(1+z_\a)^4\,n\mo\big\}\Big]+R\j \label{(2.15)}
\end{align}
if $\bowtie_j$ represents $>\,$, $\psi_j(c)=1-\chi_j(c)$ otherwise, $\ga\j$ denotes the skewness of the $j$th population, and the remainder terms $R\j$ have the properties ascribed to $R$ in section~\ref{sec:nonzero}.  

It is often unnecessary to assume, as above, that the quantile estimators $\hatt_\a\j$ are independent of one another.  To indicate why, we note that the method for deriving expansions such as \eqref{(2.8)}, \eqref{(2.10)} and \eqref{(2.12)} involves computing $P(T_c>\hatt_\a)$ by first calculating the conditional probability $P(T_c>\hatt_\a\mi\hatt_\a)$, where the independence of $T_c$ and $\hatt_\a$ is used.  Versions of this argument can be given for the case of short-range dependence among many different values of~$\hatt_\a\j$, for $1\leq j\leq p$.  However a simpler approach, giving a larger but still asymptotically negligible bound to the remainder term $O\{\ldots\}$ on the right-hand side of \eqref{(2.15)}, can be developed more simply; for brevity we do not give details~here.

Cases where the statistics are computed from weakly dependent data can be addressed using results of Hall and Wang~(2010).  That work treats instances where the variables $T_{c\j}\j$ are computed from the first $n$ components in respective data streams $\cS_j=(X_{j1},X_{j2},\ldots)$, with $X_{j1},X_{j2},\ldots$ being independent and identically distributed but correlated between streams.  As in the discussion above, since we are treating $t$-statistics then it can be assumed without loss of generality that the variables in each data stream have unit variance.  (This condition serves only to standardise scale, and in particular places the means $c\j$ on the same scale for each~$j$.)  Assuming this is the case, we shall suppose too that third moments are uniformly bounded.  Under these conditions it is shown by Hall and Wang (2010) that, provided that (a)~the correlations are bounded away from~1, (b)~the streams $\cS_1,\cS_2,\ldots$ are $k$-dependent for some fixed $k\geq1$, (c)~$z_\a$ is bounded between two constant multiples of $(\log p)\half$, (d)~$\log p=o(n)$, and (e)~for $1\leq j\leq p$ we have $0\leq c\j=c\j(n)\leq\ep\,n\mhf\,(\log p)\half$, where $\ep\ra0$ as $n\rai$; and excepting realisations that arise with probability no greater than $1-O\{p\,\exp(-C\,z_\a^2)\}$, where $C>0$; the $t$-statistics $T_{c\j}\j$ can be considered to be independent.  In particular, it can be stated that with probability $1-O\{p\,\exp(-C\,z_\a^2)\}$ there are no clusters of level exceedences caused by dependence among the data streams.  

These conditions, especially~(d), permit the dimension $p$ to be exponentially large as a function of~$n$.  Assumption~(e) is of interest; without it the result can fail and clustering can occur.  To appreciate why, consider cases where the data streams are $k$-dependent but in the degenerate sense that $\cS_{rj+1}=\ldots=\cS_{rj+k}$ for $r\geq0$.  Then, for relatively large values of $c$, the value of $T_c\j$ is well approximated by that of $c/S_j$, where $S_j^2=n\mo\,\sum_{i\leq n}\,(X_{ji}-\bX_j)^2$ is the empirical variance computed from the first $n$ data in the stream $\cS_j$.  It follows that, for any $r\geq1$, the values of $T_c^{(rj+i)}$, for $1\leq i\leq k$, are also very close to one another.  Clearly this can lead to data clustering that is not described accurately by asserting independence.

\begin{figure}[t]
\begin{center}
\vspace*{-.5cm}
\makeatletter\def\@captype{figure}\makeatother
\centering
\includegraphics[width=6in]{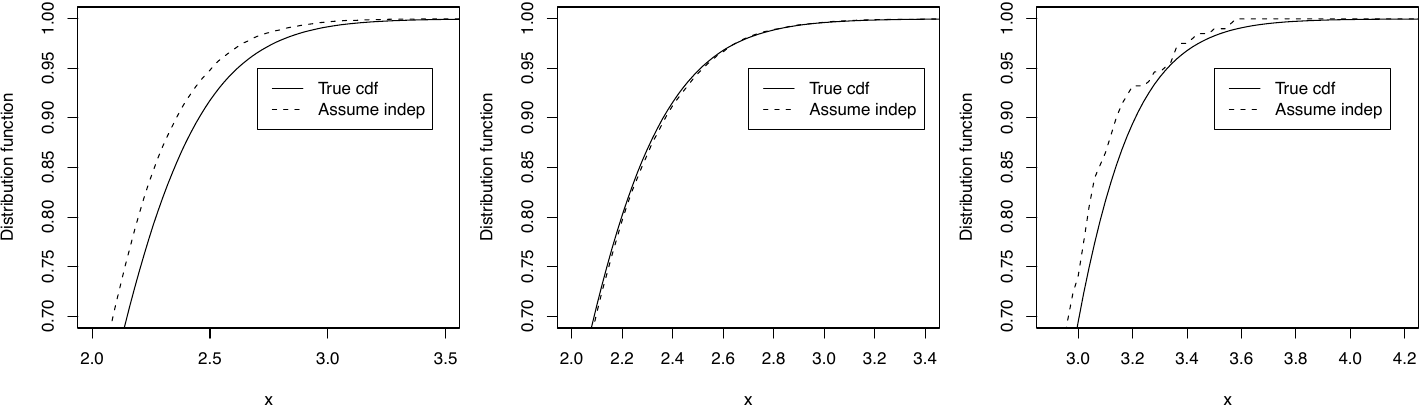}
\caption
{Comparison of the joint distribution function of $(T_0^{(1)},\ldots,T_0^{(p)})$ (denoted by ``True cdf'') with the product of the distributions of the univariate components $T_0^{(k)}$, $k=1,\ldots,p$ (denoted by ``Assume indep''), when $\epsilon_k\sim$ standardised Pareto(5,5), $n=50$ and, from left to right, $(p,\theta)=(100,0.5)$, $(p,\theta)=(100,0.2)$, $(p,\theta)=(10000,0.2)$. The vertical axis gives values of $P(T_0^{(1)}\leq x,\ldots,T_0^{(p)}\leq x)$ where $x$ is given on the horizontal axis.} \label{FigParetoDep}
\end{center}
\end{figure}

To illustrate these properties we calculated the joint distribution of $(T_0^{(1)},\ldots,T_0^{(p)})$ for short-range dependent $p$-vectors $(X_1,\ldots,X_p)$, and compared this distribution with the product of the distributions of the $p$ univariate components $T_0^{(k)}$, $k=1,\ldots,p$. For $k=1,\ldots,p$ we took $X_k=(U_k-EU_k)/\sqrt{\var U_k}$  and $U_k=\sum_{j=0}^{10} \theta^j \epsilon_{j+k}$. Here, $0<\theta<1$ is a constant and $\epsilon_1,\ldots,\epsilon_{p+10}$ denote i.i.d. random variables. Figure~\ref{FigParetoDep} depicts the resulting distribution functions for several values of $\theta$ and $p$, when the sample size $n$ was $50$ and the $\epsilon_j$s were from a standardised Pareto(5,5) distribution. We see that the independence assumption gives a good approximation to the joint cumulative distribution function, but, unsurprisingly, the approximation degrades as $\theta$ (and thus the dependence) increases. The figure also suggests that the independence approximation degrades as $p$ becomes very large ($10^5$, in this example).

\section{Application to higher criticism for detecting sparse signals in non-Gaussian noise} \label{sec:HC}
In this section we develop higher criticism methods where the critical points are based on bootstrap approximations to distributions of $t$ statistics, and show that the advantages established in section~2 for bootstrap $t$ methods carry over to sparse signal detection.  

Assume we observe $X_{1j},\ldots,X_{nj}$, for $1\leq j\leq p$, where all the observations are independent and where, for each $j$, $X_{1j},\ldots,X_{nj}$ are identically distributed. For example, in gene microarray analysis $X_{ij}$ if often used to represent the log-intensity associated with the $i$th subject and the $j$th gene, $\mu_j$ represents the mean expression level associated with the $j$th feature (i.e.~gene), and the $Z_{ij}$s represent measurement noise.  The distributions of the $X_{ij}$s are completely unknown, and we allow the distributions to differ among components. 
Let $E(X_{1j})=c^{(j)}$. The problem of signal detection is to test
\begin{equation} \label{DefineNullAlt}
\mbox{$H_0$:   all $c^{(j)}$s are zero, against\quad $H_1^{(n)}$:  a small fraction of the $c^{(j)}$s is nonzero}. 
\end{equation} 
For simplicity, in this section we assume that each $c^{(j)} \geq 0$, but a similar treatment can be given where nonzero $c^{(j)}$s have different signs.

To perform the signal detection test we use the ideas in section 2 to construct a bootstrap $t$ higher criticism statistic that can be calculated when the distribution of the data is unknown, and which is robust against heavy-tailedness of this distribution. (Higher criticism was originally suggested by Donoho and Jin (2004) in cases where the centered data have a known distribution, non-Studentised means were used, and the bootstrap was not employed.)  As in section~2.4, let $T_{c\j}\j$ be the Studentised statistic for the the $j$th component, and let $\hatt_\a\j$  be  the bootstrap estimator of the $1-\a$ quantile of the distribution of $T_0\j$, both calculated from the data $X_{1j},\ldots X_{nj}$. We suggest the following bootstrap $t$ higher criticism statistic:
\begin{equation} \label{DefineHC}
\hc_n(\alpha_0) =\max_{\alpha = i/p, \;  1 \leq i \leq \alpha_0 p }\,\{p\,\a\,(1-\a)\}\mhf\,
\sumjop\big\{I\big(T_{c\j}\j>\hatt_\a\j\big)   - \alpha  \big\}\,,  
\end{equation} 
where $\alpha_0 \in (0,1)$ is small enough for the statistic $\hc_n$ at \eqref{DefineHC} to depend only on indices $j$ for which $T_{c\j}\j$ is relatively  large.  This exploits the excellent performance of bootstrap approximation to the distribution of the Studentised mean in the tails, as exemplified by Theorems~1 and~2 in section~2, while avoiding the ``body'' of the distribution, where the bootstrap approximations are sometimes less remarkable. We reject $H_0$ if $\hc_n(\alpha_0)$ is too large.  

We could have defined the higher criticism statistic by replacing the bootstrap quantiles in definition \eqref{DefineHC} by the respective quantiles of the standard normal distribution. However, the greater accuracy of bootstrap quantiles compared to normal quantiles, established in section~2, suggest that in the higher criticism context, too, better performance can be obtained when using bootstrap quantiles.  The superiority of the bootstrap approach will be illustrated numerically in section~4.

Theorem~3 below provides upper and lower bounds for the bootstrap $t$ higher criticism statistic at \eqref{DefineHC}, under $H_0$ and $H_1^{(n)}$. We shall use these results to prove that the probabilities of type I and type II errors converge to zero as $n\to\infty$.  The standard ``test pattern'' for assessing higher criticism is a sparse signal, with the same strength at each location where it is nonzero.  It is standard to  take $c\j=0$ for all but a fraction $\eps_n$ of $j$s, and $c\j=\tau_n\,n\mhf$ elsewhere, where $\tau_n\neq 0$ is chosen to make the testing problem difficult but solvable. As usual in the higher criticism context we take 
\begin{equation} \label{Defineeps}
\eps_n =   p^{-\beta} =n^{-\beta/\theta},
\end{equation} 
where $\beta \in (0,1)$ is a fixed parameter. Among these values of $\beta$ the range $0<\beta<\thf$ is the least interesting, because there the proportion of nonzero signals is so high that it is possible to estimate the signal with reasonable accuracy, rather than just determine its existence. See Donoho and Jin (2004). Therefore we focus on the most interesting range, which is $\thf < \beta < 1$. For $\beta \in (\thf, 1)$ the most interesting values of $\tau_n$ are $\tau_n \asymp \sqrt{2 \log p}$, with $\tau_n < \sqrt{2 \log p}$.  Taking $\tau_n  = o(\sqrt{2 \log p})$ would render the two hypotheses indistinguishable, whereas taking $
\tau_n \geq  \sqrt{2 \log p}$  would render the signal relatively easy to discover, since it would imply that the means that are nonzero are of the same size as, or larger than, the largest values of the signal-free $T_{c\j}\j$s.   In light of this we consider nonzero means of size
\begin{equation} \label{Definetau}
\tau_n = \sqrt{2 r \log p}=\sqrt{2 (r/\theta)\log n}\;,
\end{equation} 
where  $0 <  r < 1$ is a fixed parameter. 

Before stating the theorem we introduce notation.  Let $L_p > 0$ be a generic multi-log term which may be different from one occurrence to the other, and is such that for any constant $c > 0$,  $L_p \cdot p^{c}\to\infty$ and $L_p \cdot p^{-c}\to 0$ as $p \to \infty$. 
We also define the ``phase function'' by 
\[
\rho_{\theta}(\beta) =  
\left\{
\begin{array}{ll} 
\Big(\sqrt{1  -  \theta} -  \sqrt{\frac{1-\theta}{2}    + \frac{1}{2}   - \beta}\, \Big)^2,    &\qquad  \thf < \beta \leq \thf + \frac{1 - \theta}{4},  \\
\beta - \thf,  &\qquad \frac{1}{2} + \frac{1 - \theta}{4} < \beta \leq \tqr, \\
(1 - \sqrt{1 - \beta})^2, &\qquad \tqr < \beta < 1. 
\end{array} 
\right. 
\]
In the $\beta$-$r$ plane we partition the region $\{\thf < \beta < 1, \rho_{\theta}(\beta) < r < 1\}$ into three subregions (i), (ii), and (iii) defined by $r < \oqr\,(1- \theta)$, $\oqr\,(1 - \theta) \leq r < \oqr$, and $\oqr < r <1$, respectively. The next theorem, derived in the longer version of this paper (Delaigle \etalc2010), provides upper and lower bounds for the bootstrap $t$ higher criticism statistic under $H_0$ and $H_1^{(n)}$, respectively.

\begin{theo} \label{lemma:HC} 
Let $p=n^{1/\theta}$, where $\theta \in (0,1)$ is fixed, and suppose that, for each $1 \leq j \leq p$,  the distribution of the respective $X$ satisfies $E(X)=0$, $E(X^2)=1$ and $E|X|^{D_2}<\infty$, where $D_2$ is chosen so large that \eqref{(2.8)} holds with $D_1>1/\theta$. Also, take $\alpha_0=n\,p^{-1} \log p$. Then

\noindent
(a) Under the null hypothesis $H_0$ in \eqref{DefineNullAlt}, there is a  constant $C > 0$ such that  
\[
P\big\{\hc_n(\alpha_0) \leq C \log p\}\to 1 \textrm{ as $n\to\infty$}\,. 
\]

\noindent
(b) Let $\beta \in (\thf, 1)$ and $r \in (0,1)$ be such that $r > \rho_{\theta}(\beta)$.  Under $H_1^{(n)}$ in \eqref{DefineNullAlt},  where $c^{(j)}$ is modeled as  in \eqref{Defineeps}--\eqref{Definetau}, we have
\[
P\{\hc_n(\alpha_0)    \geq  L_p  p^{\delta(\beta, r, \theta)}\}  \to 1 \textrm{ as $n\to\infty$}\,,
\]
where 
\[
\delta(\beta, r, \theta)  =  \left\{
\begin{array}{ll} 
 \thf -  \beta + (1 - \theta)/2 - (\sqrt{(1 - \theta)} - \sqrt{r})^2,  & {\rm if\ } (\beta,r) {\rm\ is\  in\ region \;  (i)},  \\
 r  - \beta + \thf,  & {\rm if\ } (\beta,r) {\rm\ is\  in\ region \;  (ii)},  \\
1 - \beta  -   (1 - \sqrt{r})^2,  & {\rm if\ } (\beta,r) {\rm\ is\  in\ region \;  (iii)}.    
\end{array} 
\right. 
\]

\end{theo} 
\noindent

It follows from the theorem that, if we set the test so as to reject the null hypothesis if and only if 
$\hc_n \geq  a_n$, where $a_n/\log p \to\infty$ as $n\to\infty$,  and $a_n=O(p^d)$ where $d<\delta(\beta, r, \theta)$, then as long as $r > \rho_{\theta}(\beta)$, the 
probabilities of type I and type II errors tend to zero as $n\to\infty$ (note that  $\delta(\beta, r, \theta) > 0$). 

It is also of interest to see what happens when $r < \rho_{\theta}(\beta)$, and below we treat separately the cases $r < \rho(\beta)$ and $\rho(\beta) < r < \rho_{\theta}(\beta)$, where $\rho(\beta)\equiv\rho_{1}(\beta)\geq \rho_{\theta}(\beta)$ is  the standard phase function discussed by Donoho and Jin (2004). We start with the case $r < \rho(\beta)$. There, Ingster (1999) and Donoho and Jin (2004) proved that for the sizes of $\epsilon_n$ and $\tau_n$ that we consider in \eqref{Defineeps}--\eqref{Definetau}, even when the underlying distribution of the  noise is known to be the standard normal, the sum of the probabilities of type I and type II errors of any test tends to $1$ as $n\to\infty$. See also Ingster (2001).
Since our testing problem is more difficult than this (in our case the underlying distribution of the noise is estimated from data), in this context too, asymptotically,  any test fails  if $r  < \rho(\beta)$. 

\begin{figure}
\begin{centering}
\vspace*{-.5cm}
\includegraphics[width = 6 in]{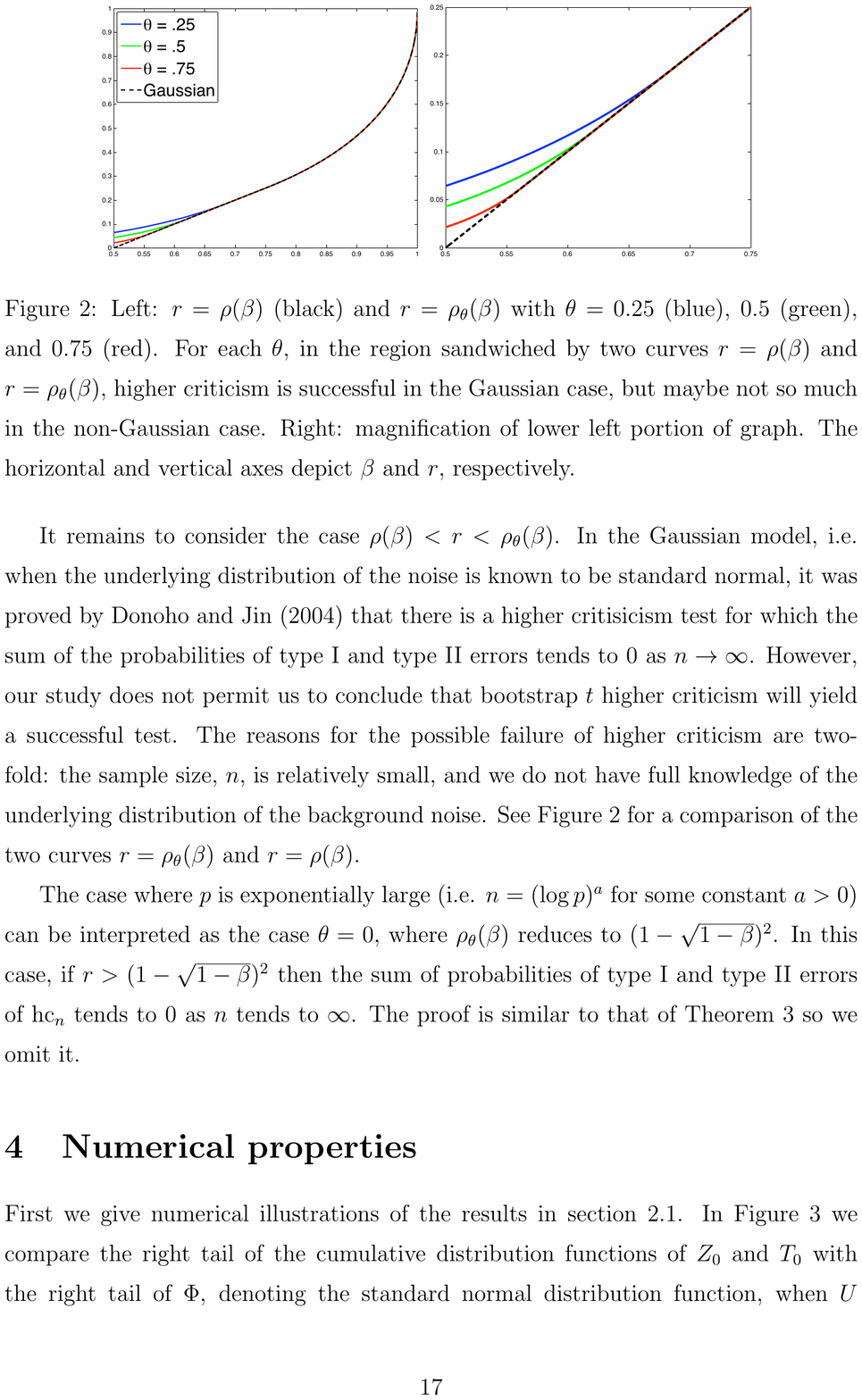}   
\caption{Left:    $r = \rho(\beta)$ (black) and $r = \rho_{\theta}(\beta)$ with $\theta = 0.25$ (blue), $0.5$ (green), and $0.75$ (red).  For each $\theta$, in the region sandwiched by two curves $r = \rho(\beta)$ and  $r = \rho_{\theta}(\beta)$,  higher criticism is successful in the Gaussian case, but maybe not so much in the non-Gaussian case.    Right: magnification of lower left portion of graph. The horizontal and vertical axes depict $\beta$ and $r$, respectively. }
\label{fig:region}
\end{centering}
\end{figure}

It remains to consider the case $\rho(\beta) < r < \rho_{\theta}(\beta)$.  In the Gaussian model, i.e. when the underlying distribution of the  noise is known to be standard normal, it was proved by Donoho and Jin (2004) that there is a higher critisicism test for which the sum of the probabilities of type I and type II errors tends to $0$ as $n\to\infty$. 
However, our study does not permit us to conclude that bootstrap $t$ higher criticism will yield a successful test. The reasons  for the possible failure of higher criticism are two-fold:  the sample size, $n$, is relatively small, and we do not have full knowledge of the underlying distribution of the background noise. See Figure \ref{fig:region} for a comparison of the two curves $r = \rho_{\theta}(\beta)$  and $r  = \rho(\beta)$.

The case where $p$ is exponentially large (i.e. $n = (\log p)^a$ for some constant $a > 0$)  can be interpreted as the case  $\theta = 0$, where  $\rho_{\theta}(\beta)$ reduces to $(1 - \sqrt{1 - \beta})^2$.  In this case,   if $r > (1 - \sqrt{1 - \beta})^2$  then  the sum of probabilities of type I and type II errors of $\hc_n$  tends to $0$ as $n$ tends to $\infty$. The proof is similar to that of Theorem \ref{lemma:HC} so we omit it.


\section{Numerical properties}\label{sec:numerical}
\begin{figure}[t]
\begin{center}
\vspace*{-.8cm}
\makeatletter\def\@captype{figure}\makeatother
\centering
\includegraphics[width=6in]{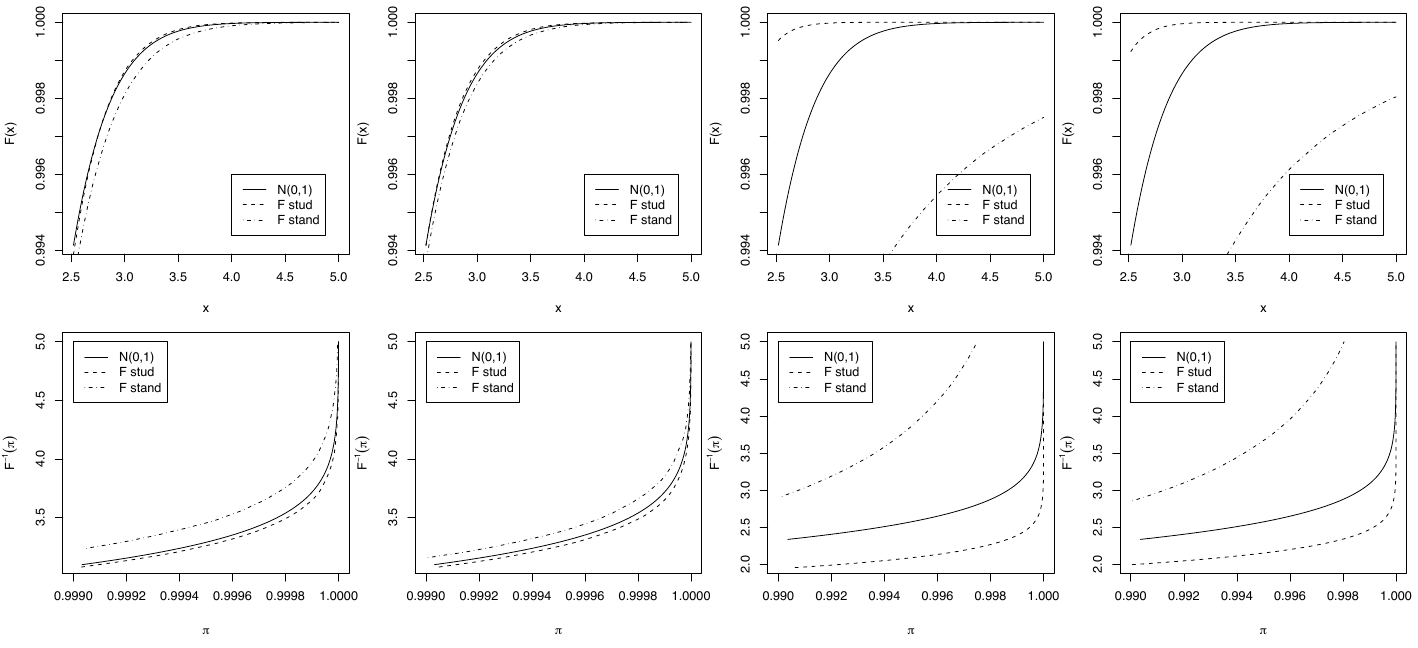} 
\caption
{Distribution function ($F$), top, and inverse distribution function ($F^{-1}$), bottom, of $T_0$ (F stud), $Z_0$ (F stand) and of a {\rm N}(0,1) when $U=N|N|$, when, from left to right $n=50$, $U=N|N|$ with $n=100$, $U=N^5|N|$ with $n=50$, $U=N^5|N|$ with $n=100$ and where $N\sim {\rm N}(0,1)$.} \label{Fig1}
\end{center}
\end{figure}

First we give numerical illustrations of the results in section \ref{sec:advantages}. In Figure \ref{Fig1} we compare the right tail of the cumulative distribution functions of $Z_0$ and $T_0$ with the right tail of $\Phi$, denoting the standard normal distribution function, when $U$ has increasingly heavy tails. We take $X=(U-EU)/(\var U)^{1/2}$ where $U=N|N|$ (moderate tails) or $N^5|N|$ (heavier tails), with $N\sim {\rm N}(0,1)$. The figure shows that $\Phi$ approximates the distribution of $T_0$ better than it approximates that of $Z_0$, and that the approximation of the normal distribution of $Z_0$ degrades as the distribution of $X$ becomes more heavy-tailed. The figure also compares the right tail of the inverse cumulative distribution functions, which shows that the normal approximation is more accurate in the tails for $T_0$ than for $Z_0$. Unsurprisingly, as the sample size increases the normal approximation for both $T_0$ and $Z_0$ becomes more accurate.

\begin{figure}[t]
\begin{center}
\vspace*{-.8cm}
\makeatletter\def\@captype{figure}\makeatother
\centering
\includegraphics[width=6in]{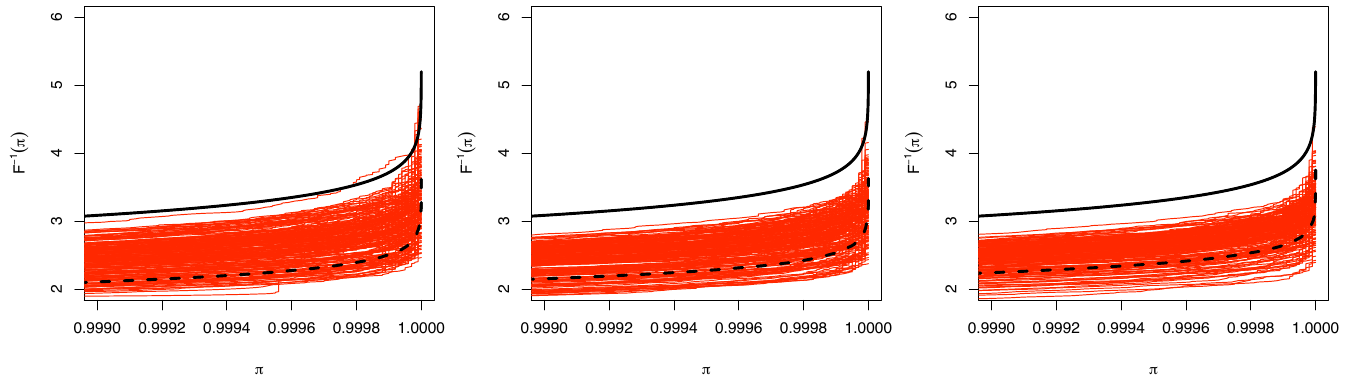} 
\caption
{Inverse ($F^{-1}$) of the distribution function of $T_0$ ( - - -), of the standard normal variable (---), and 200 bootstrap estimators of the distribution function of $T_0$ (red curves), when $X$ is a standardised ${\rm F}(5,5)$, $n=50$ (left), $n=100$ (middle), $n=250$ (right). } \label{FigFisher55}
\end{center}
\end{figure}

Next we illustrate the results in section \ref{sec:boot}. There we showed that although $T_0$ is more robust than $Z_0$ against heavy-tailedness of the distribution $F_X$ of $X$, the distribution of $T_0$ is somewhat more affected by the skewness of $F_X$. To illustrate the success of bootstrap in correcting this problem we compare the bootstrap and normal approximations for several skewed and heavy-tailed distributions. In particular, Figure \ref{FigFisher55} shows results obtained when $X=(U-EU)/(\var U)^{1/2}$, with $U\sim {\rm F}(5,5)$. Since, later in this section, we shall be more interested in approximating quantiles of the distribution of $T_0$, rather than the distribution itself, then in Figure \ref{FigFisher55} we show the right tail of the inverse cumulative distribution function of $T_0$ and 200 bootstrap estimators of this tail obtained from 200 samples of sizes $n=50$, $n=100$ or $n=250$ simulated from $F_X$. We also show the inverse cumulative distribution function of the standard normal distribution. The figure demonstrates clearly that the bootstrap approximation to the tail is more accurate than the normal approximation, and that the approximation improves as the sample size increases. We experimented with other skewed and heavy-tailed distributions, such as other F distributions and several Pareto ditributions, and reached similar conclusions. 

Note that, when implementing the bootstrap, the number $B$ of bootstrap samples has to be taken sufficiently large to obtain reasonably accurate estimators of the tails of the distribution. In general, the larger $B$, the more accurate the bootstrap approximation, but in practice we are limited by the capacity of the computer. To obtain a reasonable approximation of the tail up to the quantile $t_\alpha$, where $\alpha<\thf$,  we found that one should take $B$ no less than $100/\alpha$.

Let $\hc$ and  $\hc_{\textrm{norm}}$ denote, respectively, the theoretical and the normal versions of the higher criticism statistic, defined by the formula at the right hand side of \eqref{DefineHC}, replacing there the bootstrap quantiles $\hat t_\a\j$ by  $t_\a\j$ and $z_\alpha$, respectively, where $t_\a\j$ denote the $1-\alpha$ theoretical quantiles of $T_0^{(j)}$ and $z_\alpha$ denote the $1-\alpha$ quantile of the standard normal distribution.  To illustrate the success of bootstrap in applications of the higher criticism statistic, in our simulations we compared the statistic $\hc$ which we could use if we knew the distribution $F_X$, the bootstrap statistic $\hc_n$ defined at \eqref{DefineHC}, where the unknown quantiles $t_\a\j$ are estimated as the bootstrap quantities $\hatt_\a\j$ as discussed in the previous paragraph, and the normal version $\hc_{\textrm{norm}}$. We constructed histograms of these three versions of the higher criticism statistic, obtained from 1000 simulated values calculated under $H_0$ or an alternative hypothesis. For any of the three versions, to obtain the 1000 values we generated 1000 samples of size $n$, of $p$-vectors $(X_1,\ldots,X_p)$. We did this under $H_0$, where the mean of each $X_{j}$  was zero, and under various alternatives $H_1^{(n)}$, where we set a fraction $\eps_n$ of these means equal to $\tau_n\,n\mhf$, with $\tau_n>0$. As in section \ref{sec:HC} we took $p=n^{1/\theta}$, $\eps_n =n^{-\beta/\theta}$ and  $\tau_n=\sqrt{2r\log p}$, where we chose $\beta$ and $r$ to be on the frontier of the $r>\rho_\theta(\beta)$.

\begin{figure}[ht]
\begin{center}
\vspace*{-.75cm}
\makeatletter\def\@captype{figure}\makeatother
\centering
\includegraphics[height=7.2in]{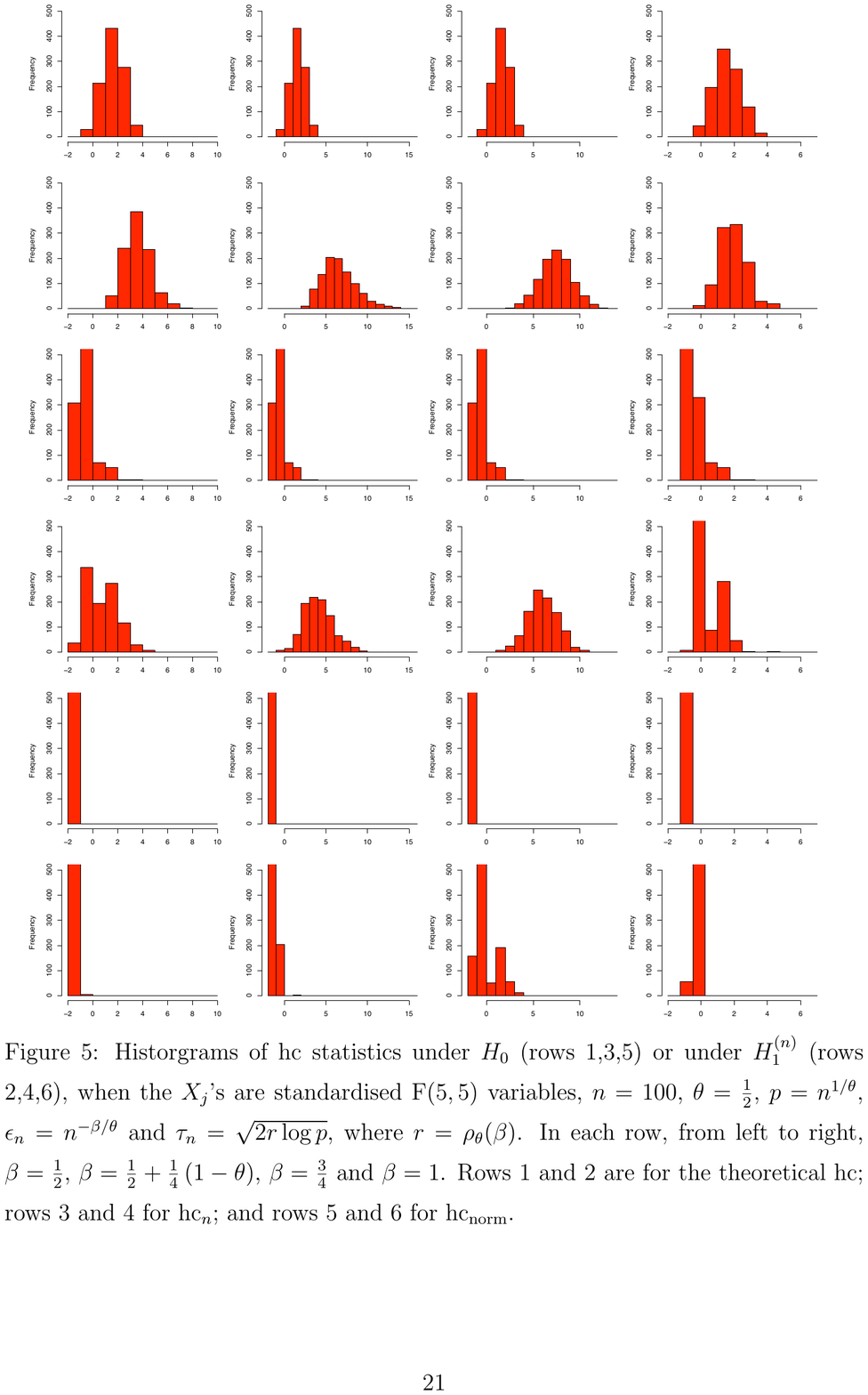}
\caption
{Historgrams of hc statistics under $H_0$ (rows 1,3,5) or under $H_1^{(n)}$ (rows 2,4,6), when the $X_j$'s are standardised ${\rm F}(5,5)$ variables, $n=100$, $\theta=\thf$, $p=n^{1/\theta}$, $\eps_n =n^{-\beta/\theta}$ and  $\tau_n=\sqrt{2r\log p}$, where $r=\rho_\theta(\beta)$. In each row, from left to right, $\beta=\thf$, $\beta=\thf+\oqr\,(1-\th)$, $\beta=\tqr$ and $\beta=1$. Rows 1 and 2 are for the theoretical $\hc$; rows 3 and 4 for $\hc_n$; and rows 5 and 6 for  $\hc_{\textrm{norm}}$.} \label{fig:HCFisher1}
\end{center}
\end{figure}

Figure \ref{fig:HCFisher1} shows the histograms under $H_0$ and under various alternatives $H_1^{(n)}$ located on the frontier ($r=\rho_\theta(\beta)$, for $\beta=\thf$, $\beta=\thf+\oqr\,(1-\th)$, $\beta=\tqr$ and $\beta=1$), when the $X_j$'s are standardised ${\rm F}(5,5)$ variables, $n=100$ and $\theta=\thf$. We can see that the histogram approximations to the density of the bootstrap  $\hc_n$ are relatively close to the histogram approximations to the density  of $\hc$. By contrast, the histograms in the case of $\hc_{\textrm{norm}}$ show that the distribution of $\hc_{\textrm{norm}}$ is a poor approximation to the distribution of $\hc$, reflecting the inaccuracy of normal quantiles as approximations to the quantiles of heavy-tailed, skewed distributions. We also see that, except when $\beta=1$,  the histograms for $\hc$ and $\hc_n$ under $H_0$ are rather well separated from those under $H_1^{(n)}$. This illustrates the potential success of higher criticism for distinguishing between $H_0$ and $H_1^{(n)}$. By contrast, this property is much less true for $\hc_{\textrm{norm}}$. 

We also compared histograms for other heavy-tailed and skewed distribution, such as the Pareto, and reached similar conclusions. Furthermore, we considered skewed but less-heavy tailed distributions, such as the chi-squared(10) distribution. There too we obtained similar results, but, while the bootstrap remained the best approximation, the normal approximation performed better than in heavy-tailed cases. We also considered values of $(\beta,r)$ further away from the frontier, and, unsurprisingly since the detection problem became easier, the histograms under $H_1^{(n)}$ became even more separated from those under $H_0$.

%

\section{Technical arguments}

\subsection{Preliminaries}
Let $T_c$ be as in \eqref{(2.11)}. Then the following result can be proved using arguments of Wang and Hall~(2009).

\begin{theoapp}  Let $B>1$ denote a constant.  Then, 
\begin{align}
{P(T_c>x)\over1-\Phi(x-c)}
&=\exp\big\{-n\mhf\,\osx\,\big(2\,x^3-3\,c\,x^2+c^3\big)\,\ga\big\}\notag\\
&\qquad\times
\Big[1+\th(c,n,x)\,\Big\{(1+|x|)\,n\mhf+(1+|x|)^4\,n\mo\Big\}\Big] \label{(5.1)}
\end{align}
as $n\rai$, where the function $\th$ is bounded in absolute value by a finite, positive constant $C_1(B)$ (depending only on $B$), uniformly in all distributions of $X$ for which $E|X|^4\leq B$, $E(X^2)=1$ and $E(X)=0$, and uniformly in $c$ and $x$ satisfying $0\leq x\leq B\,n^{1/4}$ and $0\leq c\leq u\,x$, where $0<u<1$.   
\end{theoapp}

We shall employ Theorem~A to prove the theorem below.  Details are given in a longer version of this paper (Delaigle \etalc2010).  Take $\cF$ to be any subset of the class of distributions $F$ of the random variable $X$, such that $E(|X|^{6+\ep})\leq B$ for some $\ep>0$ and a constant $1<B<\infty$, $E(X)=0$ and $E(X^2)=1$. 
Recall the definition of $T_0\as$ in \eqref{(2.6)}, let $t=t_\a$ and $t=\hatt_\a$ denote the respective solutions of $P(T_0>t)=\a$ and $P(T_0\as>t\mi\cX)=\a$, and recall that $z_\a=(1-\Phi)\mo(\a)$.  Take $\eta\in(0,\ep/\{4(6+\ep)\})$, and let $T_c$ and $\hatt_\a$ denote independent random variables with the specified marginal distributions.  

\begin{theoapp}  Let $B>1$ denote a constant.  Then, 
\begin{align}
P_F(T_c>\hatt_\a)
&=P_F(T_c>t_\a)\,\exp\big\{n\mhf\,\osx\,c\,\big(3\,z_\a^2-c^2\big)\,\ga\big\}\notag\\
&\qquad\times
\Big[1+O\big\{(1+z_\a)\,n\mhf+(1+z_\a)^4\,n\mo\big\}\Big]\notag\\
&\qquad+O\Bigg[\sum_{k=1}^3\,
P_F\bigg\{\bigg|\oon\,\sumion(1-E)\,X_i^k\bigg|>n^{-(1/4)-\eta}\bigg\}\notag\\
&\qquad\qquad\qquad\qquad\qquad
+P_F\bigg\{\bigg|\oon\,\sumion(1-E)\,X_i^4\bigg|>B\bigg\}\Bigg]
\label{(5.2)}
\end{align}
as $n\rai$, uniformly in all $F\in\cF$ and in all $c$ and $z_\a$ satisfying $0\leq z_\a\leq B\,n^{1/4}$ and $0\leq c\leq u\,z_\a$, where $0<u<1$.
\end{theoapp}

\subsection{Proof of Theorem~2}\label{sec:proofTheo2}
The following theorem can be derived from results of Adamczak~(2008).  

\begin{theoapp}If $Y_1,\ldots,Y_n$ are independent and identically distributed random variables with zero mean, unit variance and satisfying 
\begin{align}
P(|Y|>y)\leq K_1\,\exp\big(-K_2\,y^{\xi}\big)\label{(5.33)}
\end{align}
for all $y>0$, where $K_1,K_2,\xi>0$, then for each $\la>1$ there exist constants $K_3,K_4>0$, depending only on $K_1$, $K_2$, $\xi$ and $\la$, such that for all $y>0$, 
$$
P\bigg(\bigg|\sumion Y_i\bigg|>y\bigg)
\leq2\,\exp\bigg(-{y^2\over2\,\la\,n}\bigg)
+K_3\,\exp\bigg(-{y^{\xi}\over K_4}\bigg)\,.
$$
\end{theoapp}

We use Theorem~C to bound the remainder terms in Theorem~B.  If $P_F(|X|>x)\leq C_1\,\exp(-C_2\,x^{\xi_1})$ and we take $Y=(1-E)\,X^k$ for an integer $k$, then \eqref{(5.33)} holds for constants $K_1$ and $K_2$ depending on $C_1$, $C_2$ and $\xi_1$, and with $\xi=\xi_1/k$.  In particular, for all $x>0$, 
$$
P_F\bigg\{\bigg|\sumion(1-E)\,X_i^k\bigg|>x\,\big(\var X^k\big)\half\bigg\}
\leq2\,\exp\bigg(-{x^2\over2\,\la\,n}\bigg)
+K_3\,\exp\bigg(-{x^{\xi_1/k}\over K_4}\bigg)\,.
$$
Taking $k=1$, 2 or 3, and $x=x_{kn}=\const n^{(3/4)-\eta_1}$ for some $\eta_1>0$; or $k=4$ and $x=x_{kn}=\const\!$; we deduce that in each of these settings, 
\begin{align*}
P_F\bigg\{\bigg|\oon\,\sumion(1-E)\,X_i^k\bigg|>x_{nk}\bigg\}
=\begin{cases}
O\big\{\exp\big(-n^{(3\xi_1/4k)-\eta_2}\big)\big\} & \textrm{ if $k=1,2,3$}\\
O\big\{\exp\big(-n^{\xi_1/4}\big/K_5\big)\big\} & \textrm{if $k=4\,,$}
\end{cases}
\end{align*}
where $\eta_2>0$ decreases to zero as $\eta_1\downarrow0$.  Therefore the $O[\ldots]$ remainder term in \eqref{(5.2)} equals $O\{\exp(-n^{(3\xi_1/16)-\eta_2})\}$, and so Theorem~2 is implied by Theorem~B.  

\subsection{Proof of \eqref{(2.12)}} \label{proof(2.12)}
Note that, by Theorem~B in section~5.1, Theorems~1 and~2 continue to hold if we replace the left-hand sides of \eqref{(2.8)} and \eqref{(2.10)} by $P_F(T_c>\hatt_\a)$, provided we also replace the factor $\a$ on the right-hand sides by $P_F(T_c>t_\a)$.  The uniformity with which \eqref{(2.8)} and \eqref{(2.10)} hold now extends (in view of Theorem~B) to $c$ such that $0\leq c\leq u\,z_\a$ with $0<u<1$, as well as to $\a$ satisfying $0\leq z_\a\leq B\,n^{1/4}$.


\bs

\begin{center}
{{\large \bf Acknowledgement}}
\end{center}
We are grateful to Jianqing Fan and Evarist Gin\'e for helpful discussion.

\begin{center}
{{\large \bf References}}
\end{center}

\frenchspacing

\nh ADAMCZAK, R. (2008).  A tail inequality for suprema of unbounded empirical processes with applications to Markov chains.  {\sl Electron. J. Probab.} {\bf 13}, 1000-1034.  

\nh BENJAMINI, Y. AND HOCHBERG, Y. (1995).  Controlling the false discovery rate: a practical and powerful approach to multiple testing. {\sl J. Roy. Statist. Soc.} Ser.~B {\bf 57}, 289--300.


\nh BERNHARD, G., KLEIN, M. AND HOMMEL, G. (2004).  Global and multiple test procedures using ordered p-values --- a review.  {\sl Statist. Papers} {\bf 45}, 1--14.

\nh CAI, T. AND JIN, J. (2010). 
Optimal rate of convergence of estimating the null density and the proportion of non-null effects in large-scale multiple testing.  
{\sl Ann. Statist.}  {\bf 38}, 100--145.



\nh CLARKE, S. AND HALL, P. (2009).  Robustness of multiple testing procedures against dependence.  {\sl Ann. Statist.} {\bf 37}, 332--358. 

\nh DAVID, J.-P., STRODE, C., VONTAS, J., NIKOU, D., VAUGHAN, A., PIGNATELLI, P.M., LOUIS, P., HEMINGWAY, J. AND RANSON, J. (2005).  The Anopheles gambiae detoxification chip: A highly specific microarray to study metabolic-based insecticide resistance in malaria vectors.  {\sl Proc. Natl. Acad. Sci.} {\bf 102}, 4080--4084.  

\nh DELAIGLE, A. AND HALL, P. (2009).  Higher criticism in the context of unknown distribution, non-independence and classification.  In {\sl Perspectives in Mathematical Sciences I: Probability and Statistics}, 109--138.  Eds N. Sastry, M. Delampady, B. Rajeev and T.S.S.R.K. Rao. World Scientific. 

\nh DELAIGLE, A., HALL, P. AND JIN, J. (2010).  Robustness and accuracy of methods for high dimensional data analysis based on Student's $t$ statistic--long version.  

\nh DONOHO, D.L. AND JIN, J. (2004).  Higher criticism for detecting sparse heterogeneous mixtures. {\sl Ann. Statist.} {\bf 32}, 962--994.  

\nh DONOHO, D.L. AND JIN, J. (2006).  Asymptotic minimaxity of false discovery rate thresholding for sparse exponential data.  {\sl Ann. Statist.} 
{\bf 34},  2980-3018.   




\nh DUDOIT, S., SHAFFER, J.P. AND BOLDRICK, J.C. (2003).  Multiple hypothesis testing in microarray experiments.  {\sl Statist. Sci.} {\bf 18}, 73--103.


\nh FAN, J., HALL, P. AND YAO, Q. (2007).  To how many simultaneous hypothesis tests can normal, Student's $t$ or bootstrap calibration be applied? {\sl J. Amer. Statist. Assoc.} {\bf 102}, 1282--1288. 

\nh FAN, J. AND LV, J. (2008). Sure independence screening for ultrahigh dimensional feature space (with discussion). {\sl J. Roy. Statist. Soc.} Ser.~B {\bf 70}, 849--911.

\nh FINNER, H. AND ROTERS, M. (2002).  Multiple hypotheses testing and expected number of type I errors.  {\sl Ann. Statist.} {\bf 30}, 220--238.

\nh GENOVESE, C. AND WASSERMAN, L. (2004).  A stochastic process approach to false discovery control.  {\sl Ann. Statist.} {\bf 32}, 1035--1061.

\nh GIN\'E, E., G\"OTZE, F. AND MASON, D.M. (1997). When is the Student $t$-statistic asymptotically standard normal? {\sl Ann. Probab.} {\bf 25}, 1514--1531.  IS IT 2007 OR 1997?

\nh HALL, P. (1990).  On the relative performance of bootstrap and Edgeworth approximations of a distribution function. {\sl J. Multivariate Anal.} {\bf 35}, 108--129.

\nh HALL, P. AND WANG, Q. (2004). Exact convergence rate and leading term in central limit theorem for Student's $t$ statistic. {\sl Ann. Probab.} {\bf 32}, 1419--1437.

\nh HALL, P. AND WANG, Q. (2010).  Strong approximations of level exceedences related to multiple hypothesis testing. {\sl Bernoulli}, to appear.

\nh INGSTER, Yu. I. (1999).  Minimax detection of a signal for $l\sp n$-balls. {\sl Math. Methods Statist.} {\bf 7}, 401--428.  

\nh INGSTER, Yu. I. (2001).  Adaptive detection of a signal of growing dimension. I. Meeting on Mathematical Statistics. {\sl Math. Methods Statist.} {\sl 10}, 395--421.





\nh JIN, J. (2007). Proportion of nonzero normal means:  universal oracle equivalences and uniformly consistent estimators.    {\sl J. Roy. Statist. Soc.} Ser.~B {\bf 70}, 461--493. 


\nh JIN, J. AND CAI. T. (2007).  Estimating the null and the proportion of  non-null effects in large-scale multiple comparisons.  {\sl J. Amer. Statist. Assoc.} {\bf 102}, 496--506. 


\nh KESSELMAN, H.J., CRIBBIE, R. AND HOLLAND, B. (2002).  Controlling the rate of Type I error over a large set of statistical tests.  {\sl Brit. J. Math. Statist. Psych.} {\bf 55}, 27--39.

\nh KULINSKAYA, E. (2009).  On fuzzy familywise error rate and false discovery rate procedures for discrete distributions.  {\sl Biometrika} {\bf 96}, 201--211.  

\nh LANG, T.A. AND SECIC, M. (1997).  {\sl How to Report Statistics in Medicine: Annotated Guidelines for Authors.}  American College of Physicians, Philadelphia.  

\nh LEHMANN, E.L., ROMANO, J.P. AND SHAFFER, J.P. (2005). On optimality of stepdown and stepup multiple test procedures.  {\sl Ann. Statist.} {\bf 33}, 1084--1108.

\nh LINNIK, JU. V. (1961).  Limit theorems for sums of independent quantities, taking large deviations into account. I. {\sl Teor. Verojatnost. i Primenen} {\bf 7}, 145--163.  




\nh PETROV, V.V. (1975).  {\sl Sums of Independent Random Variables}.  Springer, Berlin.  

\nh PIGEOT, I. (2000).  Basic concepts of multiple tests --- A survey.  {\sl Statist. Papers} {\bf 41}, 3--36.

\nh SARKAR, S.K. (2006).  False discovery and false nondiscovery rates in single-step multiple testing procedures.  {\sl Ann. Statist.} {\bf 34}, 394--415.


\nh SHAO, Q.-M. (1999).  A Cram\'er type large deviation result for Student's $t$-statistic.  {\sl J. Theoret. Probab.} {\bf 12}, 385--398.

\nh STUDENT (1908).  The probable error of a mean.  {\sl Biometrika} {\bf 6}, 1--25.

\nh TAKADA, T., HASEGAWA, T., OGURA, H., TANAKA, M., YAMADA, H., KOMURA, H. AND ISHII, Y. (2001).  
Statistical filter for multiple test noise on fMRI.  {\sl Systems and Computers in Japan} {\bf 32}, 16--24.  

\nh TAMHANE, A.C. AND DUNNETT, C.W. (1999).  Stepwise multiple test procedures with biometric applications.  {\sl J. Statist. Plann. Inf.} {\bf 82}, 55--68.  

\nh WANG, Q. AND HALL, P. (2009). Relative errors in central limit theorems for Student's $t$ statistic, with applications. {\sl Statist. Sinica} {\bf 19}, 343--354.  

\nh WU, W.B. (2008).  On false discovery control under dependence.  {\sl Ann. Statist.} {\bf 36},    364--380.





\nonfrenchspacing

\newpage

\appendix
\label{page:app}

\pagestyle{headings}
\section{{\bf PAGES \pageref{page:app}--\pageref{page:append}: NOT-FOR-PUBLICATION APPENDIX}}

\subsection{Proof of Theorem~B}

\ni{\sl Step~1: Expansions of $t_\a$ and~$\hatt_\a$.}  
The main results here are \eqref{(5.5)} and \eqref{(5.7)}.  To derive them, take $W\as$ to have the distribution of $(X_i\as-\bX)/S$, where $S$ is as in \eqref{(2.1)}, and, for $k=3$ and~4, put 
$$
\hga_k=E\big(W\as{}^k\bigmi\cX\big)={1\over n\,S^k}\,\sumion(X_i-\bX)^k\,,
$$
where $W\as{}^k=(W\as)^k$.  Letting $c=0$ in Theorem~A, and taking $X$ there to have the distribution of $W\as$ conditional on $\cX$, we deduce that if $B>1$ is given, 
\begin{align}
{P_F(T_0\as>x\mi\cX)\over1-\Phi(x)}
&=\exp\big(-n\mhf\,\otd\,x^3\,\hga\big)\notag\\
&\qquad\times
\Big[1+\Th_1(n,x)\,\Big\{(1+|x|)\,n\mhf+(1+|x|)^4\,n\mo\Big\}\Big]\,,\label{(5.3)}
\end{align}
where $\hga=\hga_3$ and:
\\[.5cm]
\medskip
\hspace*{.5cm}
\begin{minipage}{.85\linewidth}
\setlength{\baselineskip}{0.8\baselineskip}
the random function $\Th_1(n,x)$ satisfies $|\Th_1(n,x)|\leq C_1(B)$ (where $C_1(B)$ 
is the same constant introduced in Theorem~A) uniformly in datasets
$\cX$ for which $S>\thf$ and $\hga_4\leq B$, and uniformly also in $x$ satisfying
$0\leq x\leq B\,n^{1/4}$.
\end{minipage}
\begin{minipage}{.1\linewidth}
\vspace*{-1cm}
\begin{equation}
\label{(5.4)}
\end{equation}
\end{minipage}

Properties \eqref{(5.3)} and \eqref{(5.4)} imply that
$\hatt_\a$ satisfies:
\begin{align}
\hatt_\a=z_\a\,\Big[1-\otd\,\hga\,n\mhf\,z_\a
+\Th_2(n,\a)\,\big\{(1+z_\a)\mo\,n\mhf+(1+z_\a)^2\,n\mo\big\}\Big]\,,\label{(5.5)}
\end{align}
where $z=z_\a$ is the solution of $1-\Phi(z_\a)=\a$ and, in the case $j=2$:
\\[.5cm]
\medskip
\hspace*{.5cm}
\begin{minipage}{.85\linewidth}
\setlength{\baselineskip}{0.8\baselineskip}
the random function $\Th_j(n,\a)$ satisfies $|\Th_j(n,\a)|\leq C_j(B)$ (where $C_j(B)$
is a finite, positive constant) uniformly in datasets $\cX$ for which $S>\thf$
and $\hga_4\leq B$, and uniformly also in $\a$ satisfying $\thf\leq 1-\a\leq1-\Phi(B\,n^{1/4})$
\end{minipage}
\begin{minipage}{.1\linewidth}
\vspace*{-1cm}
\begin{equation}
\label{(5.6)}
\end{equation}
\end{minipage}

Analogously, Theorem~A implies that $t_\a$ satisfies:
\begin{align}
t_\a=z_\a\,\Big[1-\otd\,\ga\,n\mhf\,z_\a
+\th(n,\a)\,\big\{(1+z_\a)\mo\,n\mhf+(1+z_\a)^2\,n\mo\big\}\Big]\,,\label{(5.7)}
\end{align}
where $z=z_\a$ is the solution of $1-\Phi(z_\a)=\a$ and:
\\[.5cm]
\medskip
\hspace*{.5cm}
\begin{minipage}{.85\linewidth}
\setlength{\baselineskip}{0.8\baselineskip}
the function $\th(n,\a)$ satisfies $|\th(n,\a)|\leq C_2(B)$ (with $C_2(B)$ denoting
a finite, positive constant) uniformly in distributions of $X$ for which
$E(X)=0$, $E(X^2)=1$ and $E(X^4)\leq B$, and uniformly also in $\a$ satisfying $0\leq z_\a\leq B\,n^{1/4}$.
\end{minipage}
\begin{minipage}{.1\linewidth}
\vspace*{-1cm}
\begin{equation}
\label{(5.8)}
\end{equation}
\end{minipage}

The derivations of the pairs of properties \eqref{(5.5)} and \eqref{(5.6)}, and \eqref{(5.7)} and \eqref{(5.8)}, are similar.  For example, suppose that if $\hatt_\a$ is given by \eqref{(5.5)} rather than by $P(T_0\as>\hatt_\a\mi\cX)=\a$, and that the function $\Th_2$ in \eqref{(5.5)} is open to choice except that it should satisfy~\eqref{(5.6)}.  If we define $\rho(z)=z\,\{1-\Phi(z)\}/\phi(z)=1-z\mt+3\,z^{-4}-\ldots$, then by \eqref{(5.3)}, \eqref{(5.5)} and~\eqref{(5.6)}, 
\begin{align}
P_F(T_0\as>\hatt_\a\mi\cX)
&=\{1-\Phi(\hatt_\a)\}\,\exp\big(-n\mhf\,\otd\,\hatt_\a^3\,\hga\big)\notag\\
&\qquad\times
\Big[1+\Th_1(n,\hatt_\a)\,\Big\{(1+|\hatt_\a|)\,n\mhf+(1+|\hatt_\a|)^4\,n\mo\Big\}\Big]\notag\\
&=(2\pi)\mhf\,
\exp\big\{-\thf\,z_\a^2\,\big(1-\ttd\,\hga\,n\mhf\,z_\a\big)
-n\mhf\,\otd\,z_\a^3\,\hga\big\}\notag\\
&\qquad\times
z_\a\mo\,\rho(z_\a)\,\Big[1+\Th_3(n,z_\a)\,\Big\{(1+z_\a)\,n\mhf+(1+z_\a)^4\,n\mo\Big\}\Big]\notag\\
&=\{1-\Phi(z_\a)\}\,\Big[1+\Th_3(n,z_\a)\,\Big\{(1+z_\a)\,n\mhf
+(1+z_\a)^4\,n\mo\Big\}\Big]\,,\notag\\
&\label{(5.9)}
\end{align}
where $\Th_3$ satisfies~\eqref{(5.6)}.  By judicious choice of $\Th_2$, satisfying \eqref{(5.6)}, we can ensure that $\Th_3$ in \eqref{(5.9)} vanishes, up to the level of discreteness of the conditional distribution function of $T_0\as$.  In this case the right-hand side of \eqref{(5.9)} equals simply $1-\Phi(z_\a)=\a$, so that $\hatt_\a$ indeed has the intended property, i.e.~$P(T_0\as>\hatt_\a\mi\cX)=\a$.  

\ni{\sl Step~2: Expansions of the difference between $\hatt_\a$ and $t_\a$.}  
The main results here are \eqref{(5.12)} and \eqref{(5.13)}.  To obtain them, first combine \eqref{(5.5)} and \eqref{(5.7)} to deduce that:
\begin{align}
\hatt_\a-t_\a
=\otd\,z_\a^2\,(\ga-\hga)\,n\mhf
+\Th_4(n,\a)\,\big\{n\mhf+(1+z_\a)^3\,n\mo\big\}\,,\label{(5.10)}
\end{align}
where, for $j=4$:
\\[.5cm]
\medskip
\hspace*{.5cm}
\begin{minipage}{.85\linewidth}
\setlength{\baselineskip}{0.8\baselineskip}
the random function $\Th_j(n,\a)$ satisfies $|\Th_j(n,\a)|\leq C_j(B)$ (with $C_j(B)$
denoting a finite, positive constant) uniformly in datasets $\cX$ for which
$S>\thf$ and $\hga_4\leq B$; uniformly in distributions of $X$ for which $E(X)=0$,
$E(X^2)=1$ and $E(X^4)\leq B$; and uniformly also in $\a$ satisfying $0\leq z_\a\leq$
$B\,n^{1/4}$.
\end{minipage}
\begin{minipage}{.1\linewidth}
\vspace*{-1cm}
\begin{equation}
\label{(5.11)}
\end{equation}
\end{minipage}

Using \eqref{(5.7)}, \eqref{(5.8)}, \eqref{(5.10)} and \eqref{(5.11)} we deduce that:
\begin{align*}
t_\a^2\,(\hatt_\a-t_\a)
&=\otd\,z_\a^4\,n\mhf\,(\ga-\hga)\,
+\Th_5(n,\a)\,\big\{(1+z_\a)^2\,n\mhf+(1+z_\a)^5\,n\mo\big\}\,,\notag\\
t_\a\,(\hatt_\a-t_\a)^2
&=\onn\,z_\a^5\,n\mo\,(\ga-\hga)^2
+\Th_6(n,\a)\,\big\{(1+z_\a)^3\,n\mo+(1+z_\a)^6\,n^{-3/2}\big\}
\end{align*}
and $(\hatt_\a-t_\a)^3=\Th_7(n,\a)\,(1+z_\a)^6\,n^{-3/2}\leq\Th_8(n,\a)\,(1+z_\a)^4\,n\mo$, where $\Th_5,\ldots,\Th_9$ (the latter appearing below) satisfy \eqref{(5.11)}.  Therefore, 
\begin{align}
\hatt_\a^3-t_\a^3
&=3\,t_\a^2\,(\hatt_\a-t_\a)+3\,t_\a\,(\hatt_\a-t_\a)^2+(\hatt_\a-t_\a)^3\notag\\
&=z_\a^4\,n\mhf\,(\ga-\hga)
+\Th_9(n,\a)\,\big\{(1+z_\a)^2\,n\mhf+(1+z_\a)^5\,n\mo\big\}.\label{(5.12)}
\end{align}

Similarly, using \eqref{(5.7)} and \eqref{(5.10)}, 
\begin{align}
(\hatt_\a-c)^2-(t_\a-c)^2
&=2\,(t_\a-c)\,(\hatt_\a-t_\a)+(\hatt_\a-t_\a)^2\notag\\
&=\ttd\,(z_\a-c)\,z_\a^2\,(\ga-\hga)\,n\mhf\notag\\
&\qquad+\Th_{10}(c,n,\a)\,\big\{(1+z_\a)\,n\mhf
+(1+z_\a)^4\,n\mo\big\}\,,\label{(5.13)}
\end{align}
where, for $j\geq10$:
\\[.5cm]
\medskip
\hspace*{.5cm}
\begin{minipage}{.85\linewidth}
\setlength{\baselineskip}{0.8\baselineskip}
the random function $\Th_j(c,n,\a)$ satisfies $|\Th_j(c,n,\a)|\leq C_j(B)$ (with
$C_j(B)$ denoting a finite, positive constant) uniformly in datasets $\cX$ for
which $S>\thf$ and $\hga_4\leq B$, uniformly in distributions of $X$ for which
$E(X)=0$, $E(X^2)=1$ and $E(X^4)\leq B$, and uniformly also in $c$ such
that $0\leq c\leq u\,z_\a$ where $0<u<1$, and in $\a$ such that $0\leq z_\a\leq B\,n^{1/4}$.
\end{minipage}
\begin{minipage}{.1\linewidth}
\vspace*{-1cm}
\begin{equation}
\label{(5.14)}
\end{equation}
\end{minipage}
\bs

\ni{\sl Step~3: Initial expansion of $P(T_c>\hatt_\a)$.}
To derive \eqref{(5.20)}, the main result in this step, note that by \eqref{(5.5)}--\eqref{(5.7)} and \eqref{(5.13)}, 
\begin{align}
1-\Phi(\hatt_\a-c)
&=(z_\a-c)\mo\,\rho(z_\a-c)\,(2\pi)\mhf\,\exp\big\{-\thf\,(\hatt_\a-c)^2\big\}\notag\\
&\qquad\times
\Big[1+\Th_{11}(c,n,\a)\,\big\{(1+z_\a)\,n\mhf+(1+z_\a)^2\,n\mo\big\}\Big]\notag\\
&=(z_\a-c)\mo\,\rho(z_\a-c)\,(2\pi)\mhf\notag\\
&\qquad\times
\exp\Big\{-\thf\,(t_\a-c)^2-\otd\,(z_\a-c)\,z_\a^2\,(\ga-\hga)\,n\mhf\Big\}\notag\\
&\qquad\times
\Big[1+\Th_{12}(c,n,\a)\,\big\{(1+z_\a)\,n\mhf+(1+z_\a)^4\,n\mo\big\}\Big]\notag\\
&=\{1-\Phi(t_\a-c)\}\,\exp\big\{-\otd\,(z_\a-c)\,z_\a^2\,(\ga-\hga)\,n\mhf\big\}\notag\\
&\qquad\times
\Big[1+\Th_{13}(c,n,\a)\,\big\{(1+z_\a)\,n\mhf+(1+z_\a)^4\,n\mo\big\}\Big]\,.
\label{(5.15)}
\end{align}

If $T_c$ is statistically independent of $\hatt_\a$ then, by \eqref{(5.1)}, \eqref{(5.12)} and \eqref{(5.13)} (the latter with $c=0$), 

\begin{align}
&\!\!\!\!\!\!\!\!\!\!\!\!
{P_F(T_c>\hatt_\a\mi\hatt_\a)\over1-\Phi(\hatt_\a-c)}\notag\\
&=\exp\big\{-n\mhf\,\osx\,\big(2\,\hatt_\a^3-3\,c\,\hatt_\a^2+c^3\big)\,\ga\big\}\notag\\
&\qquad\times
\Big[1+\Th_{14}(c,n,\a)\,\Big\{(1+z_\a)\,n\mhf+(1+z_\a)^4\,n\mo\Big\}\Big]\notag\\
&=\exp\big\{-n\mhf\,\osx\,\big(2\,t_\a^3-3\,c\,t_\a^2+c^3\big)\,\ga\big\}\notag\\
&\qquad\times
\exp\Big[-n\mo\,\otd\,\ga\,(\ga-\hga)\,
\big\{z_\a^4-c\,z_\a^3\big\}\Big]\notag\\
&\qquad\times
\Big[1+\Th_{15}(c,n,\a)\,\Big\{(1+z_\a)\,n\mhf+(1+z_\a)^4\,n\mo\Big\}\Big]\notag\\
&={P_F(T_c>t_\a)\over1-\Phi(t_\a-c)}\,
\Big[1+\Th_{16}(c,n,\a)\,\Big\{(1+z_\a)\,n\mhf+(1+z_\a)^4\,n\mo\Big\}\Big]\,.
\label{(5.16)}
\end{align}

Combining \eqref{(5.15)} and \eqref{(5.16)} we deduce that:
\begin{align}
P_F(T_c>\hatt_\a\mi\hatt_\a) 
&=P_F(T_c>t_\a)\cdot\exp\big\{-\otd\,(z_\a-c)\,z_\a^2\,(\ga-\hga)\,n\mhf\big\}\notag\\
&\qquad\times
\Big[1+\Th_{17}(c,n,\a)\,\big\{(1+z_\a)\,n\mhf+(1+z_\a)^4\,n\mo\big\}\Big]\,.
\label{(5.17)}
\end{align}

Reflecting \eqref{(5.14)}, let $\cG_1(B)$ denote the class of distribution functions $F$ of $X$ such that $E(X)=0$, $E(X^2)=1$ and $E(X^4)\leq B$; write $P_F$ for probability measure when $\cX$ is drawn from the population with distribution function $F\in\cG_1$; let $\cD$ denote any given event, shortly to be defined concisely; let $\cE(B)$ be the intersection of $\cD$ and the events $S>\thf$ and $\hga_4\leq B$; and write $\tcE(B)$ for the complement of $\cE(B)$.  In view of~\eqref{(5.17)}, 
\begin{align}
P_F(T_c>\hatt_\a)
&=P_F(T_c>t_\a)\cdot E\Big[\exp\Big\{n\mhf\,\otd\,(z_\a-c)\,z_\a^2\,(\hga-\ga)\Big\}\,I\{\cE(B)\}\Big]\notag\\
&\qquad\times
\Big[1+O\big\{(1+z_\a)\,n\mhf+(1+z_\a)^4\,n\mo\big\}\Big]\notag\\
&\qquad+O\big[P_F\big\{\tcE(B)\big\}\big]\,,
\label{(5.20)}
\end{align}
uniformly in the following sense:
\\[.5cm]
\medskip
\hspace*{.5cm}
\begin{minipage}{.85\linewidth}
\setlength{\baselineskip}{0.8\baselineskip}
uniformly in $F\in\cG_1(B)$, in $c$ such that $0\leq c\leq u \,z_\a$, where $0<u<1$, and
in $\a$ such that $0\leq z_\a\leq B\,n^{1/4}$.
\end{minipage}
\begin{minipage}{.1\linewidth}
\vspace*{-1cm}
\begin{equation}
\label{(5.21)}
\end{equation}
\end{minipage}

\bs

\ni{\sl Step~4: Simplification of right-hand side of~\eqref{(5.20)}.}
Here we derive a simple formula, \eqref{(5.32)}, for the expectation on the right-hand side of~\eqref{(5.20)}.  That result, when combined with \eqref{(5.20)} and \eqref{(5.21)}, leads quickly to Theorem~B.

Put $\De_k=n\mo\,\sumi(X_i^k-EX^k)$, write $\cD_k$ to denote the event that $|\De_k|\leq C_3\,n^{-(1/4)-\eta}$ where $C_3>0$ and $\eta\in(0,\thf)$, and put $\cD=\cD_1\cap\cD_2\cap\cD_3$.  Observe that  
\begin{align}
\hga=\big(\ga+\De_3-3\,\De_1\,\De_2-3\Delta_1+2\,\De_1^3\big)\big/
\big(1+\De_2-\De_1^2\big)\,,\label{(5.22)}
\end{align}
From this property it can be proved that if $|\ga|\leq B$ and $C_3$ is sufficiently small, depending only on $B$, then $|\hga-\ga|\leq n^{-(1/4)-\eta}$ whenever $\cD$ holds.  Therefore if $\cD$ holds, and $0\leq c\leq u\,z_\a$ for $0<u<1$, and $0\leq z_\a\leq B\,n^{1/4}$, then 
$$
n\mhf\,|z_\a-c|\,z_\a^2\,|\hga-\ga|
\leq B^3\,n^{-\eta}\,.
$$
In these circumstances, defining $\De=n\mhf\,\otd\,(z_\a-c)\,z_\a^2\,(\hga-\ga)$, we have:
\begin{align}
\bigg|e^\De-\sum_{j=0}^r\,{\De^j\over j!}\bigg|\,I(\cD)
\leq\big(B^3\,n^{-\eta}\big)^{r+1}\,\exp\big(B^3\,n^{-\eta}\big)\,.\label{(5.23)}
\end{align}

Note too that if $E(X^6)\leq B$ then 
\\[.5cm]
\medskip
\hspace*{.5cm}
\begin{minipage}{.85\linewidth}
\setlength{\baselineskip}{0.8\baselineskip}
$E(\De_{k_1}^{r_1}\,\De_{k_2}^{r_2})\leq C_4(B)\,n\mo$ whenever $k_1$ and $k_2$ take values in the set
$\{1,2,3\}$, $r_1$ and $r_2$ are nonnegative, and $r_1+r_2=1$ or~2.
\end{minipage}
\begin{minipage}{.1\linewidth}
\vspace*{-1cm}
\begin{equation}
\label{(5.24)}
\end{equation}
\end{minipage}

Also, in the same context as \eqref{(5.24)}, if $r_1+r_2=2$ then 
\begin{align}
E\big\{|\De_{k_1}^{r_1}\,\De_{k_2}^{r_2}|\,I\big(\tcD\big)\big\}
\leq E\big(|\De_{k_1}^{r_1}\,\De_{k_2}^{r_2}|\big)
\leq\big\{E\big(\De_{k_1}^{2r_1}\big)\,
E\big(\De_{k_2}^{2r_1}\big)\big\}\half\leq  C_4(B)\,n\mo\,;\label{(5.25)}
\end{align}
and if $r_1=1$ and $r_2=0$, and $\eta$ is sufficiently small, 
\begin{align}
E\big\{|\De_{k_1}^{r_1}\,\De_{k_2}^{r_2}|\,I\big(\tcD\big)\big\}
\leq\big\{E\De_{k_1}^{2r_1}\,P\big(\tcD\big)\big\}\half
&\leq C_5(B,\eta)\,\big(n\mo\,n^{-(1/2)-\ze}\big)\half\notag\\
&=C_5(B,\eta)\,n^{-(3/4)-(\ze/2)}\label{(5.26)}
\end{align}
where $\ze=\ze(\eta)>0$.  In deriving \eqref{(5.26)} we used the fact that $P(\tcD)\leq P(\tcD_1)+P(\tcD_2)+P(\tcD_3)$, and that, by Markov's inequality (employing the fact that $E|X|^{6+\ep}\ab<\infty$ and choosing $\eta<\ep/\{4(3+\ep)\}$), 
\begin{align*}
P\big(\tcD_k\big)&\leq\big(C_3\,n^{-(1/4)-\eta}\big)^{-\{2+(\ep/3)\}}\,
E\big(|\De_k|^{2+(\ep/3)}\big)\\
&\leq C_6(B,\eta)\,n^{(1/2)+2\eta+(\ep/12)+(\eta\ep/3)-\{2+(\ep/3)\}/2}
\leq C_6(B,\eta)\,n^{-(1/2)-\ze}
\end{align*}
for $k=1,2,3$, where $\ze>0$.  Therefore, 
\begin{align}
P\big(\tcD\big)\leq3\,C_6(B,\eta)\,n^{-(1/2)-\ze}\,.\label{(5.27)}
\end{align}
If $r_1+r_2+r_3\geq3$ then an argument similar to that leading to \eqref{(5.27)} shows that 
\begin{align}
E\big\{\big|\De_1^{r_1}\,\De_2^{r_2}\,\De_3^{r_3}\big|\,I(\cD)\big\}
\leq C_7(B,\eta)\,\big(n\mhf\big)^2\,\big(n^{-(1/4)-\eta}\big)^{r_1+r_2+r_3-2}\,.
\label{(5.28)}
\end{align}
Combining \eqref{(5.24)}, \eqref{(5.25)}, \eqref{(5.26)} and \eqref{(5.28)}; using Taylor expansion to derive approximations to $\hga-\ga$, starting from \eqref{(5.22)}; noting the definition of $\De$ given in the previous paragraph; and observing that $n\mhf\,|z_\a-c|\,z_\a^2\leq B^3\,n^{1/4}$ if $0\leq c\leq u\,z_\a$ with $0<u<1$, and $0\leq z_\a\leq B\,n^{1/4}$; we deduce that:
\begin{align}
\big|E\big\{\De^j\,I(\cD)\big\}\big|
&\leq
\begin{cases}
C_8(B,j)\,n^{1/4}\,n\mo&\textrm{if $j=1$}\\
C_8(B,j)\,\big(n^{1/4}\big)^j\,n\mo\,\big(n^{-(1/4)-\eta}\big)^{j-2}& \textrm{if $j\geq2$}
\end{cases}
\notag\\
&\leq C_8(B,j)\,n\mhf\,.\label{(5.29)}
\end{align}

Using \eqref{(5.23)}, \eqref{(5.27)} and \eqref{(5.29)}, and choosing $r$ to be the least integer such that $(r+1)\,\eta\geq\thf$, we deduce that:
\begin{align}
E\Big[\exp\Big\{n\mhf\,\otd\,(z_\a-c)\,z_\a^2\,(\hga-\ga)\Big\}\,I(\cD)\Big]
=1+O\big(n\mhf\big)\,,\label{(5.30)}
\end{align}
uniformly in the following sense:
\\[.5cm]
\medskip
\hspace*{.5cm}
\begin{minipage}{.85\linewidth}
\setlength{\baselineskip}{0.8\baselineskip}
uniformly in $F\in\cG_2(B)$, in $c$ such that $0\leq c\leq u\,z_\a$ with $0<u<1$, and
in $\a$ such that $0\leq z_\a\leq B\,n^{1/4}$,
\end{minipage}
\begin{minipage}{.1\linewidth}
\vspace*{-1cm}
\begin{equation}
\label{(5.31)}
\end{equation}
\end{minipage}

where $\cG_2(B)$ denotes the intersection of $\cG_1(B)$ (defined at \eqref{(5.21)}) with the class of distributions of $X$ such that $E(|X|^{6+\ep})\leq B$.  

An argument almost identical to that leading to \eqref{(5.30)} and \eqref{(5.31)} shows that the same pair of results holds if we replace $\cD$ by the event $\cD_1$ that $S>\thf$ and $\hga_4\leq B$.  The only change needed is the observation that, since $F\in\cG_1(B)$ entails $E(|X|^{6+\ep})\leq B$,  $P(\tcD_1)$ is uniformly bounded above by a constant multiple of~$n\mhf$.  This follows from the fact that, if $Y_1,Y_2,\ldots$ are random variables satisfying $E|Y|^{6+\ep}<\infty$, then $P\{|\sum_{i\leq n}\,(1-E)\,Y_i^4|>n\}\leq\const\,n^{-(1/2)-(\ep/4)}$.  Therefore, in the argument in \eqref{(5.26)} we can replace the bound $\const n^{-(1/2)-\ze}$ to $P(\tcD)$ by the bound $\const n^{-(1/2)-(\ep/4)}$ to~$P(\tcD_1)$.  This means that \eqref{(5.30)} holds if we replace $\cD$ there by the event $\cD\cap\cD_1$, i.e.~the event $\cE(B)$ introduced just above~\eqref{(5.20)}.  That~is, 
\begin{align}
E\Big[\exp\Big\{n\mhf\,\otd\,(z_\a-c)\,z_\a^2\,(\hga-\ga)\Big\}\,I\{\cE(B)\}\Big]
=1+O\big(n\mhf\big)\,,\label{(5.32)}
\end{align}
uniformly in the sense of~\eqref{(5.31)}.  

Together, \eqref{(5.20)}, \eqref{(5.21)} and \eqref{(5.32)} imply that \eqref{(5.2)} holds uniformly in $F\in\cF$, in $c$ such that $0\leq c\leq u\,z_\a$, with $0<u<1$ and in $\a$ such that $0\leq z_\a\leq B\,n^{1/4}$, completing the proof of Theorem~B.

\subsection{Proof of Theorem \ref{lemma:HC}} \label{sec:proofJiashun}
Throughout this proof we use the notation $\hc_n^* = \hc_n(\alpha_0^*)$, where $\alpha_0^* = n\,p^{-1} \log p$ denotes the value of $\alpha_0$ stated in the theorem.  
Also, for two positive sequences $a_n$ and $b_n$, we write $a_n \lesssim b_n$ when $\limsup_{n \goto \infty} (a_n/b_n) \leq 1$.  We use the equivalent notation $b_n \gtrsim a_n$.

Fix $\alpha \in (0,1)$. Let  $G_p(\alpha)  = p^{-1} \sum_{j = 1}^p  I\big(T_{c\j}\j>\hatt_\a\j\big)$, and $\hc_{n, \alpha} =\sqrt{p}  \{  \alpha (1 - \alpha) \}^{-1/2} \{G_p(\alpha) - \alpha\}$.  We have $\hc_n^* = \max_{\alpha = i/p,  1 \leq i \leq \alpha_0^* p} \hc_{n, \alpha}$. We introduce a non-stochastic counterpart 
$$\widetilde{\hc}_n^* = \max_{\alpha = i/p, 1 \leq i \leq \alpha_0^* p}  \widetilde{\hc}_{n,\alpha}$$
of $\hc_n^*$, where $\widetilde{\hc}_{n,\alpha} = \sqrt{p}  \{  \alpha (1 - \alpha) \}^{-1/2} \{\bar{G}_p(\alpha) - \alpha\}$ and $\bar{G}_p(\alpha)  =  p^{-1} \sum_{j = 1}^p  P\big(T_{c\j}\j>\hatt_\a\j\big)$.  Note that $\bar{G}_p(\alpha) = E\{G_p(\alpha)\}$.

The keys for the proofs are:  
\begin{description} 
\item [(A)]    There is a  constant $C > 0$ such that 
\begin{align*} 
\lim_{n \goto \infty} P\Big\{|\hc_n^*  - \widetilde{\hc}_n^*| \leq C \log p   \,\Big\}  = 1,  &\qquad  \mbox{under  $H_0$}, \\
\lim_{n \goto \infty} P\Big\{|\hc_n^*  - \widetilde{\hc}_n^*| \leq C \log p\, \sqrt{1 + \widetilde{\hc}_n^*}\,\Big\}  =1,  &\qquad \mbox{under $H_1^{(n)}$}. 
\end{align*} 
\item [(B)]  Under $H_0$,  there is a constant $C > 0$ such that $\widetilde{\hc}_n^*  \leq C \log p$ for sufficiently large $n$. 
\item [(C)]  Under $H_1^{(n)}$,  $\widetilde{\hc}_n^*   =  L_p  p^{\delta(\beta, r, \theta)}$. 
\end{description} 
Combining {\bf(A)}--{\bf(B)}, there exit constants $C_1 > 0$ and $C_2 > 0$ such that  $\widetilde{\hc}_n^* \leq C_1 \log p$ and $P\{  |\hc_{n}^* - \widetilde{\hc}_n^*|   \leq C_2 \log p  \} = 1 + o(1)$. Therefore, 
\[
P \Big\{ \hc_n^*  \leq  (C_1 + C_2) \log p\Big\}  \geq  P\Big\{| \hc_n^*  - \widetilde{\hc}_n^*|  \leq C_2  \log p  \Big\}   =   1 + o(1), 
\]
and part (a) of Theorem \ref{lemma:HC} follows.   Combining {\bf(A)} and {\bf(C)} gives that
\[
P\Big\{\hc_n^*  \geq  L_p p^{\delta(\beta, r, \theta)}\Big\}   \geq  P\Big\{\hc_n^*  \geq \widetilde{\hc}_n^*  -  C \log p \,\sqrt{1 + \widetilde{\hc}_n^*}   \Big\}\to 1 \textrm{ as $n\to\infty$},  
\]
and part (b) of the theorem follows.  Note that $C$ and $L_p$ may stand for  different quantities in different occurrence. 

We now show {{\bf(A)}--{\bf(C)}.    Below, whenever we refer to $\alpha$,  we assume that  
$p^{-1} \leq \alpha \leq \alpha_0^*$.  By definition, 
$\bar{G}_p(\alpha) = p^{-1} \sum_{j =1}^p P(T_{c^{(j)}}^{(j)}  > \hat{t}_{\alpha}^{(j)})$,  
where the fraction of $c^{(j)} = 0$ is $1$ under the null and $(1 - \eps_n)$ under the alternative.    Using Theorem 1 and noting  that  $O(n^{-D_1}) = o(1/p)$ and  that   $z_{\alpha} \leq O(\sqrt{\log p})$ in \eqref{(2.8)}, we have
\begin{equation} \label{cf1.1}
P(T_{c^{(j)}}^{(j)}  > \hat{t}_{\alpha}^{(j)})   = \alpha \{1 + O(\sqrt{\log p}/\sqrt{n})\} + o(1/p), \qquad  \mbox{when $c^{(j)} = 0$}. 
\end{equation} 
 It follows that both under the null and under the alternative, 
\begin{equation} \label{galpha}
(1 - \eps_n) \alpha \lesssim \bar{G}_p(\alpha) \lesssim  (1 - \eps_n) \alpha + \eps_n.    
\end{equation}  
As a result,   uniformly in $\alpha \in [1/p, \alpha_0^*]$, 
\begin{equation} \label{galpha1}
\alpha = o(1), \qquad \bar{G}_p(\alpha) = o(1), \qquad p\, \bar{G}_p(\alpha) \gtrsim p\, \alpha \geq 1. 
\end{equation} 

Consider {\bf(A)}.   Note that for any integer $N \geq 1$ and  any  positive sequences $a_i$ and $b_i$,      $\max_{1 \leq i \leq N} \{ a_i b_i \} \leq \max_{1 \leq i \leq N} \{ a_i \} \cdot  \max_{1 \leq i \leq N}  \{b_i\}$. By the definition of $\hc_n^*$ and $\widetilde{\hc}_n^*$,   
\[
|\hc_n^* - \widetilde{\hc}_n^*| \leq \max_{\alpha = i/p:  1 \leq i \leq \alpha_0^* p}  \frac{\sqrt{p}\  |G_p(\alpha) -  \bar{G}_p(\alpha)|}{\sqrt{\alpha (1 - \alpha)}} \leq I \cdot II, 
\]
where $I$ is stochastic and $II$ is deterministic, and 
\[
I  =  \max_{\alpha = i/p:  1 \leq i \leq \alpha_0^* p}  \frac{\sqrt{p}\,  |G_p(\alpha) -  \bar{G}_p(\alpha)|}{\sqrt{\bar{G}_p(\alpha) (1 - \bar{G}_p(\alpha))}}, 
\qquad II =   \max_{\alpha = i/p:  1 \leq i \leq \alpha_0^* p}  \frac{ |\bar{G}_p(\alpha) (1 -   \bar{G}_p(\alpha))|}{\sqrt{\alpha  (1 - \alpha)}}.  
\]
To show {\bf(A)}, it is sufficient to show that both under the null and the alternative, 
\begin{equation} \label{apf1}
P(I \geq C \log p)  = o(1),  
\end{equation} 
and that 
\begin{equation} \label{apfa1}
II \lesssim 1 \;\;\;    \mbox{under $H_0$}, \qquad II     \lesssim  \sqrt{1 + | \widetilde{\hc}_n^*|}  \;\;\;  \mbox{under $H_1^{(n)}$}. 
\end{equation} 

Consider  \eqref{apf1}.     Note that 
\begin{equation} \label{apf1.1} 
P(I > C \log p) \leq \sum_{\alpha = i/p, 1 \leq i \leq \alpha_0^* p}  P\biggl\{\frac{\sqrt{p}\,  |G_p(\alpha) -  \bar{G}_p(\alpha)|}{\sqrt{\bar{G}_p(\alpha) (1 - \bar{G}_p(\alpha))}} >  C \log(p)\biggr\}.  
\end{equation} 
For each $\alpha$, 
applying Bennett's inequality [Shorack and Wellner (1986) page 851]  with $X_j = I(T_{c^{(j)}}^{(j)} > \hatt_{\alpha}^{(j)})  - P(T_{c^{(j)}}^{(j)} > \hatt_{\alpha}^{(j)})$  
and $\lambda = C \log(p) \sqrt{\bar{G}_p(\alpha) (1 - \bar{G}_p(\alpha))}$,      
\begin{align} 
P\Bigg\{ \frac{\sqrt{p} \, |G_p(\alpha) -  \bar{G}_p(\alpha)|}{\sqrt{\bar{G}_p(\alpha) \{1 - \bar{G}_p(\alpha)\}}}
 C \log p\Bigg\}  \notag
\equiv &\ P\big\{\sqrt{p}\, |G_p(\alpha) - \bar{G}_p(\alpha) | \geq \lambda\big\}  \\
\leq &\ 2\, \mathrm{exp}  \biggl\{ -    \frac{\lambda^2}{2\sigma^2} \psi\Big(  \frac{2 \lambda}{\sigma^2 \sqrt{p}}\Big)\biggr\}, \label{apf1.2}
\end{align} 
where     $\psi(\lambda) = (2/\lambda^2) \{(1 + \lambda) \log(1 + \lambda) - 1\}$ is  monotonely decreasing in $\lambda$ and  satisfies  $\lambda^2 \psi(\lambda) \sim 2 \lambda  \log \lambda$ for large $\lambda$, and  
$\sigma^2$ is the average variance of $X_j$: 
\[
\sigma^2 =   \frac{1}{p} \sum_{j =1}^p \biggl[P(T_{c^{(j)}}^{(j)} > \hatt_{\alpha}^{(j)}) -  \big\{P\big(T_{c^{(j)}}^{(j)} > \hatt_{\alpha}^{(j)}\big)\big\}^2 \biggr].  
\]

On one hand,  recall that there is at least a fraction $(1 - \eps_n)$ of $c^{(j)}$s that are $0$, and that  when $c^{(j)} = 0$,  $P(T_{c^{(j)}}^{(j)} > \hatt_{\alpha}^{(j)}) \sim \alpha$.   We see that
\begin{equation} \label{apf1.21a} 
\sqrt{p}\, \sigma \gtrsim  \sqrt{p} \,\sqrt{\alpha}  \geq 1.    
\end{equation} 
On the other hand,  by Schwartz inequality,  
\[ 
	\sigma^2  \leq  \frac{1}{p}\sum_{j =1}^p  P\big(T_{c^{(j)}}^{(j)} > \hatt_{\alpha}^{(j)}\big)  -  \Big\{\frac{1}{p}\sum_{j =1}^p  P\big(T_{c^{(j)}}^{(j)} > \hatt_{\alpha}^{(j)}\big)\Big\}^2  =   \bar{G}_p(\alpha) \{1 - \bar{G}_p(\alpha)\}. 
\]
It follows from the definition of $\lambda$ that  
\begin{equation} \label{apf1.21b} 
\frac{\lambda}{\sigma} \geq C \log p. 
\end{equation} 
Recalling that $\psi$ is monotonely decreasing, and that 
$\lambda^2 \psi(\lambda) \sim 2 \lambda \log(\lambda)$ for large $\lambda$,  it follows from \eqref{apf1.21a}--\eqref{apf1.21b}  that 
\begin{equation} \label{apf1.3}
2\mathrm{exp}  \biggl\{ -    \frac{\lambda^2}{2\sigma^2} \psi\Big(  \frac{2 \lambda}{\sigma^2 \sqrt{p}}\Big)\biggr\} 
\leq 2\mathrm{exp}  \biggl\{ -    \frac{\lambda^2}{2\sigma^2} \psi\Big(  \frac{C \lambda}{\sigma}\Big)\biggr\}  
\leq 2 \mathrm{exp}  \biggl[ -  C \Big\{\frac{\lambda}{\sigma}\log\big(\frac{\lambda}{\sigma}\big) \Big\}\biggr], 
\end{equation} 
where $C >  0$ is a generic constant. 
Note that the last term in \eqref{apf1.3}   $= o(1/p)$.   Combining 
\eqref{apf1.1}--\eqref{apf1.2} and  \eqref{apf1.21b}--(\ref{apf1.3})  gives   \eqref{apf1}. 

It remains to prove \eqref{apfa1}.    Recall that  $\bar{G}_p(\alpha) = o(1)$. 
By the definition of $II$,  
\begin{equation} \label{apf1.4}
II = \max_{\alpha=i/p, 1 \leq i \leq \alpha_0^* p}  \sqrt{\frac{\bar{G}_p(\alpha) (1 - \bar{G}_p(\alpha))}{\alpha (1 - \alpha)}}\lesssim   \max_{\alpha=i/p, 1 \leq i \leq \alpha_0^* p}  \sqrt{\frac{\bar{G}_p(\alpha)}{\alpha}}.
\end{equation} 
Under the null,  by Theorem 1,   
 $\bar{G}_p(\alpha) / \alpha = 1 + O(\sqrt{\log p}/\sqrt{n}) + o(1)$,
which gives the first assertion in \eqref{apfa1}.   For the second assertion, write 
$$\frac{\bar{G}_p(\alpha)}{\alpha}    =  1 +   \frac{\bar{G}_p(\alpha) - \alpha}{\alpha}  = 1 +  
\widetilde{\hc}_{n, \alpha} \frac{\sqrt{\alpha(1 - \alpha)}}{\sqrt{p}\, \alpha}\ \cdot$$  
Noting  that $\sqrt{\alpha (1 - \alpha)} /(\sqrt{p}\, \alpha) \leq  1$,  it follows that 
\begin{equation}  \label{apf1.5}
\frac{\bar{G}_p(\alpha)}{\alpha} \leq 1 +  |\widetilde{\hc}_{n, \alpha}|.
\end{equation} 
Combining \eqref{apf1.4} and \eqref{apf1.5} gives the second claim of \eqref{apfa1}.

Consider {\bf(B)}.     In this case, the null hypothesis is true and all $c^{(j)}$s equal $0$.   By the definition and \eqref{cf1.1}--\eqref{galpha1},   
\begin{equation} \label{cf1.10}
|\widetilde{\hc}_{n, \alpha}| \leq   \frac{\sqrt{p}\, |\bar{G}_p(\alpha) - \alpha|}{\sqrt{\alpha(1 - \alpha)}}  \lesssim  C \sqrt{\alpha \cdot  \log  p \cdot  (p /n)} + o(1).   
\end{equation} 
Recalling  that  $\alpha \leq    \alpha_0^*$ with $\alpha_0^* = (n/p) \log p$, $\widetilde{\hc}_{n, \alpha} \leq C \log p$ and the claim follows.

Consider {\bf(C)}.  In this case, the alternative hypothesis is true, and a fraction $(1 - \eps_n)$ of $c^{(j)}$s is $0$, with the remaining of them equal to $\tau_n$.   
Using (2.13),  where $c = \tau_n$, 
\begin{equation} \label{cf1.2}
P( T_{c^{(j)}}  > \hat{t}_{\alpha}^{(j)}) = (1 + o(1)) \cdot  P( T_{c^{(j)}}  > t_{\alpha}) + o(1/p) =    \bar{\Phi}(z_{\alpha} - \tau_n)  + o(1/p), 
\end{equation} 
where $\bar{\Phi} = 1 - \Phi$ is the survival function of a ${\rm N}(0,1)$. 
Combining \eqref{cf1.1} and  \eqref{cf1.2},  
\[
\bar{G}_p(\alpha) = (1 - \eps_n) \alpha(1 + O(\sqrt{\log p}/\sqrt{n}) + \eps_n L_p \bar{\Phi}(z_{\alpha} - \tau_n) + o(1/p), 
\]
and 
it follows from direct calculations that 
\begin{align} \label{cf1.3}
\widetilde{\hc}_{n, \alpha}    
=& \frac{\sqrt{p}\, [\bar{G}_p(\alpha)  - \alpha]}{\sqrt{\alpha (1 - \alpha)}}  \notag  \\
& = \frac{\sqrt{p} [\eps_n    L_p \bar{\Phi}(z_{\alpha} - \tau_n)  +  (1 - \eps_n) \alpha (1 + O(\sqrt{\log(p)/n})) - \alpha + o(1/p)] }{\sqrt{\alpha(1 - \alpha)}}  \notag\\ 
 & =  \frac{ L_p  \sqrt{p} \,\eps_n \bar{\Phi}(z_{\alpha} - \tau_n)}{\sqrt{\alpha(1 - \alpha)}}   -  \sqrt{p} \eps_n \sqrt{\frac{\alpha}{(1 - \alpha)}}   + O(\sqrt{p \log p\, \alpha/n}) + o(1).   
\end{align} 
Recall that  $\alpha \leq \alpha_n^*  =n\, p^{-1}\, \log p$. First,  $\sqrt{p} \eps_n \sqrt{\alpha/(1 - \alpha)} \lesssim \eps_n \sqrt{p \alpha_0^*}  \leq   \eps_n \sqrt{n \log(p)}$. This   equals $L_p p^{\theta/2 -  \beta} = o(1)$ because $\theta < 1$ and $\beta > \thf$. Second,  $\sqrt{p \log p\cdot \alpha/n} \leq  \sqrt{p \log(p) \alpha_0^*}  \leq  \log p$. 
Inserting these into \eqref{cf1.3} gives 
\[
\widetilde{\hc}_{n, \alpha}    = \frac{ L_p  \sqrt{p}\, \eps_n \bar{\Phi}(z_{\alpha} - \tau_n) }{\sqrt{\alpha(1 - \alpha)}}   + L_p, 
\] 
and so 
\begin{equation} \label{cf1.45}
\widetilde{\hc}_n^* = III + L_p, \qquad \mbox{where}  \; III =  \max_{\alpha = i/p, 1 \leq i \leq \alpha_0^* p}  \frac{L_p  \sqrt{p}\, \eps_n \bar{\Phi}(z_{\alpha} - \tau_n)}{\sqrt{\alpha(1 - \alpha)}}. 
\end{equation} 

We  now re-parametrize with $z_{\alpha}$ as
\[ 
z_{\alpha} =  \sqrt{2 q \log p}  \equiv s_n(q), \textrm{ where $q>0$, } \mbox{so that } \qquad \alpha = \bar{\Phi}\{s_n(q)\}.  
\]
By Mill's ratio,  we have $\bar{\Phi}(s_n(q))   = L_p p^{-q}$. Recall that $1/p \leq \alpha \leq \alpha_0^*$, where $\alpha_0^* = L_p p^{\theta - 1}$.    We deduce that the range of possible values for the parameter $q$ runs from $(1 - \theta)$ to $1$ (with lower order terms   neglected). It follows from elementary calculus that 
\begin{equation} \label{cf1.5}
III  =  \max_{(1 - \theta) \leq 	 q \leq 1}   \frac{L_p \sqrt{p}\, \eps_n \bar{\Phi}\{s_n(q) - \tau_n\}}{\sqrt{\bar{\Phi}\{s_n(q)\} [1 - \bar{\Phi}\{s_n(q)\}]}} = L_p \cdot  \max_{(1 - \theta) \leq 	 q \leq 1}   \frac{ \sqrt{p}\, \eps_n \bar{\Phi}\{s_n(q) - \tau_n\}}{p^{-q/2}}.   
\end{equation} 
Moreover, by Mill's ratio, 
\begin{equation} \label{cf1.6}
\sqrt{p}\, \eps_n \bar{\Phi}\{s_n(q) - \tau_n\}  =   
L_p  \cdot p^{\pi(q, \beta, r)}, 
\end{equation} 
where 
\[ 
\pi(q, \beta, r) = \left\{   
\begin{array}{ll} 
 \thf  - \beta,     &\qquad   0 < q < r,      \\ 
 \thf  - \beta  - (\sqrt{q} - \sqrt{r})^2,      &\qquad    r < q < 1.   
\end{array} 
\right. 
\]
Inserting \eqref{cf1.6} into  \eqref{cf1.5}  gives 
\begin{equation} \label{cpf2}
III   = L_p \cdot \max_{(1  - \theta) \leq q \leq 1}  p^{\pi(q;  \beta, r) + q/2}. 
\end{equation} 

We now analyze $\pi(q;  \beta, r) + q/2$ as a function of $q \in (0,1]$. 
In region (i),  $4r  \leq (1 - \theta)$, and  $\pi(q; \beta, r) + q/2$ is monotonely 
decreasing in $[(1 - \theta),1]$.  Therefore, the maximizing value of $q$ is  $(1 - \theta)$,  at which  
$\pi(q; \beta, r) + q/2 = \thf  - \beta + (1 - \theta)/2  - \{\sqrt{(1 - \theta)} - \sqrt{r}\}^2$. 
In region (ii),  $(1 - \theta) < 4r   \leq  1$.   As $q$ ranges between $(1 - \theta)$ and $1$, $\pi(q; \beta, r) + q/2$ first  monotonely increases and reaches the maximum at $q = 4r$, then monotonely decreases. The maximum of  $ \pi(q; \beta, r) + q/2$ is then $r  - \beta  + \thf$. 
In region (iii), $4r > 1$, and  
$\pi(q; \beta, r) + q/2$ is monotonely 
increasing in $[(1 - \theta),1]$. 
The maximizing value of $q$  is  $1$,  at which  $\pi(q; \beta, r) + q/2 =   1 - \beta -  (1 - \sqrt{r})^2$.   
Combining these  with \eqref{cpf2}  and (\ref{cf1.45})  gives the claim.    \qed 

\bs

\begin{center}
{{\large \bf References}}
\end{center}

\nh SHORACK, G.R. AND WELLNER, J.A. (1986). Empirical Process with Application to Statistics. {\sl John Wiley \& Sons, NY}.

\frenchspacing
\nonfrenchspacing

\label{page:append}

\end{document}